\documentstyle[natbib,emulateapj]{article}

\def\etal   {{et~al.\/}}

\slugcomment{To Appear in PASP, June 2004}

\begin{document}

\title{SparsePak: A Formatted Fiber Field Unit for The WIYN
Telescope Bench Spectrograph. I. Design, Construction, and Calibration}

\author{Matthew A. Bershady} \affil{Astronomy Dept., U. Wisconsin -
Madison, 475 N. Charter St., Madison, WI 53706; mab@astro.wisc.edu}

\author{David R. Andersen\altaffilmark{1,2}} \affil{Dept. of Astronomy
\& Astrophysics, Penn State University, University Park, PA 16802}

\author{Justin Harker\altaffilmark{3}} \affil{Astronomy Dept.,
U. Wisconsin - Madison, 475 N. Charter St., Madison, WI 53706}

\author{Larry W. Ramsey}
\affil{Dept. of Astronomy \& Astrophysics, Penn State University,
University Park, PA 16802}

\author{Marc A. W. Verheijen\altaffilmark{4,5}} \affil{Astronomy
Dept., U. Wisconsin - Madison, 475 N. Charter St., Madison, WI 53706}

\altaffiltext{1}{CIC Scholar, UW-Madison.}

\altaffiltext{2}{Curent Address: Max Planck Institut f\"{u}r
Astronomie, K\"{o}nigstuhl 17, 69117 Heidelberg, Germany}

\altaffiltext{3}{Current Address: UCO/Lick Observatory, Astronomy
Department, University of California, Santa Cruz, CA 95064}

\altaffiltext{4}{McKinney Fellow, UW-Madison}

\altaffiltext{5}{Current Address: Astrophysikalisches Institut Potsdam,
An der Sternwarte 16, 14482, Potsdam, Germany}

\smallskip

\smallskip
\begin{abstract}

We describe the design and construction of a formatted fiber
field-unit, SparsePak, and characterize its optical and astrometric
performance. This array is optimized for spectroscopy of low-surface
brightness, extended sources in the visible and
near-infrared. SparsePak contains 82, 4.7$^{\prime\prime}$ fibers
subtending an area of 72$^{\prime\prime}\times$71$^{\prime\prime}$ in
the telescope focal plane, and feeds the WIYN Bench
spectrograph. Together, these instruments are capable of achieving
spectral resolutions of $\lambda/\Delta\lambda\sim 20,000$ and an
area--solid-angle product of $\sim 140$ arcsec$^2$ m$^2$ per
fiber. Laboratory measurements of SparsePak lead to several important
conclusions on the design of fiber termination and cable curvature to
minimize focal ratio degradation. SparsePak itself has throughput
$>80$\% redwards of 5200\AA\, and 90-92\% in the red. Fed at f/6.3,
the cable delivers an output 90\% encircled energy at nearly
f/5.2. This has implications for performance gains if the WIYN Bench
Spectrograph had a faster collimator. Our approach to integral-field
spectroscopy yields an instrument which is simple and inexpensive to
build, yet yields the highest area--solid-angle product per spectrum
of any system in existence. An Appendix details the fabrication
process in sufficient detail for others to repeat. SparsePak was
funded by the National Science Foundation and the University of
Wisconsin-Madison Graduate School, and is now publicly available on
the WIYN Telescope through the National Optical Astronomical
Observatories.

\end{abstract}


\section{INTRODUCTION}

Observational astronomy consists of obtaining subsets of a fundamental
data hyper-cube of the apparent distribution of photons in angle$^2$
on the sky $\times$ wavelength $\times$ time $\times$
polarization. Information-gathering systems (``instruments'') are
designed to make science-driven trades on the range and sampling of
each of these dimensions. Here we describe an instrument optimized for
the study of the stellar and ionized gas kinematics in disks of nearby
and distant galaxies. Such studies require bi-dimensional spectroscopy
at medium spectral resolution ($5000<R<20,000$, where
$R\equiv\lambda/\Delta\lambda$) of extended sources over a relatively
narrow range of wavelength, e.g., $\sim$600 spectral channels, with no
consideration of time-sampling or polarization.

What is paramount for our application is the ability to gather
sufficient signal at low light levels and medium spectral
resolution. Etendue at constant spectral resolution and sampling is
the relevant figure of merit (etendue is the product of area, solid
angle, and system throughput).  To characterize the light-gathering
aperture alone in what follows we refer to ``grasp'' (the area
$\times$ solid angle product).

Field-of-view and spatial resolution are also of merit, but
of secondary importance. For our particular application, we require
the number of spatial resolution elements be sufficient to resolve
galaxy disks out to 3-4 scale lengths at several points per
scale-length -- about 15 points across a diameter. The matching of
overall scale is dictated, then, by the telescope and fiber size
needed to reach the required spectral resolution and etendue with an
affordable spectrograph.  That is, the galaxies are to be chosen to
fit the instrument.


Bacon \etal\ (1995) and Ren \& Allington-Smith (2002) discuss the
trades between spatial and spectral dynamic range and sampling for a
variety of spectroscopic instruments -- all fundamentally limited by
the two-dimensional sampling geometry of their detector focal planes.
An evaluation of this discussion yields that integral-field
spectroscopy (IFS) and Fabry-Perot (FP) imaging are the preferred
methods for bi-dimensional spectroscopy at low light levels. Each has
its merits. Relative to imaging FP systems, IFS systems trade spatial
resolution and coverage for spectral resolution and spectral coverage.
For emission-line studies in need of high spatial resolution FP are
optimal. For lower spatial resolution, particularly at high spectral
resolution where the FP ``bull's eye'' is small, IFS is particularly
competitive. (The ``bull's eye'' is the angular diameter in which the
central wavelength shifts by less than the spectral resolution.) For
absorption line studies where many spectral channels are needed, IFS
is superior. This condition becomes even more pronounced if spectral
resolution and etendue are valued more highly than spatial resolution,
as we do here.



Within the realm of ``integral-field'' spectroscopy there is still a
wide range of spatial and spectral sampling (see Ren \&
Allington-Smith 2002). Our interest is in fiber-fed or image-slicing
systems which reformat the telescope focal plane (i.e., into a slit),
optimizing for spectral sampling while still retaining bi-dimensional
spatial coverage (cf. SAURON, Bacon \etal\ 2001a, whic is optimized
for spatial sampling at the sacrifice of spectral sampling).
While image slicers minimize entropy increases, fibers relax the
restriction on the spatial sampling. In some situation, the entropy
increase (i.e., information loss) from fibers via focal-ratio
degradation (FRD; see Angel \etal\ 1977; Barden, Ramsey \& Truax,
1981; Clayton 1989; or Carrasco \& Parry 1994) may be more than
compensated by the increased flexibility in how the fibers sample the
telescope focal plane and are mapped into the dispersive optical
system. 

For example, fibers offer the advantage of formatted sampling, whereby
fibers intentionally do not sample an integral area, but instead some
structured, coherent, yet dispersed pattern.  While this
``patterning'' has not been used in fixed-bundle arrays, there are
clear advantages to this approach. For an extended object, all spatial
elements do not carry equal scientific weight, and there is a trade
between coverage and sampling. Fibers allow these trades to be
fine-tuned. In the specific example of galaxy kinematic studies, one
may choose to allocate fibers to sample both major and minor axes, or
perhaps uniformly cover a larger physical area with sparse sampling
such that lower surface-brightness outer regions are still sampled
more frequently than the inner, higher surface-brightness cores. The
latter is the approach we have adopted here.

These considerations led us to our design of a {\it formatted} fiber
array for the WIYN Bench Spectrograph -- the first of its kind. It is
a simple, inexpensive instrument with dramatic gains in information
gathering power for a broad range of scientific programs which require
bi-dimensional, medium-resolution spectroscopy at low
surface-brightness levels. The SparsePak design is in an orthogonal
direction to most 8m-class IFU instruments which strive to maximize
spatial resolution at high cost and complexity and at the loss of
medium spectral resolution for background limited observations.

Our strategy was to take advantage of an existing spectrograph capable
of echelle resolutions with a very long slit (equivalent to about 12
arcmin) to optimize the study of galaxy kinematics at low surface
brightness. The result of this mating of fiber cable to spectrograph
yields a bi-dimensional spectroscopic system capable of achieving
spectral resolutions and grasp comparable to the best systems on {\it
any} telescope. The specific grasp of SparsePak -- the grasp per
spatial resolution element -- is the highest of any spectroscopic
instrument (see Figure 1). It is this singular attribute that allows
SparsePak to be used effectively at low surface-brightness and medium
spectral resolutions. This fiber array was completed on March 10,
2001, installed in May 2001, and successfully commissioned over the
following month.

In a series of two papers we describe the design, construction, and
performance of this array and its associated spectrograph.  We start
in \S 2 with a basic description of SparsePak and the spectrograph it
feeds, as well as a comparison of SparsePak to existing fiber-arrays
on WIYN and other telescopes. In \S 3 of this paper we present the
design goals and constraints which led to the SparsePak
formatted-array design. A synopsis of SparsePak's technical attributes
are presented in \S 4. The construction of single-fiber reference
cables are also documented. This cable provides a benchmark for
evaluating the laboratory-measured performance of SparsePak.  Sections
5 and 6 contain the astrometric and optical properties of the array,
as measured in our lab. In \S 7 is a summary of our results.  An
appendix contains a technical description of the fabrication process
in sufficient detail for others to repeat the process.

In Paper II (Bershady \etal\ 2003) we establish the on-sky performance
of the array and spectrograph in terms of throughput, spectral
resolution and scattered-light; we demonstrate the ability to perform
precision sky-subtraction and spectrophotometry; and we present
examples of commissioning science data which highlight the
capabilities for which SparsePak was designed.

\section{BASIC DESCRIPTION}

SparsePak is a formatted, fiber-optic field unit which pipes light
from the f/6.3 Nasmyth imaging (``WIYN'') port to the WIYN Bench
Spectrograph (Barden \etal\ 1994). The plate-scale at this port is
9.374 arcsec/mm. The highly-polished SparsePak fibers have no
fore-optics nor anti-reflection coatings.  The WIYN port has no
corrector nor ADC. Hence there are only the three reflective telescope
surfaces up-stream of the fibers. The cable transmittance was measured
in the lab to be near 85\% at 6500\AA.

At the WIYN port, each of SparsePak's 82 active fibers has an active
diameter of 4.687$^{\prime\prime}$. SparsePak's grasp is $\sim$11,200
m$^2$ arcsec$^2$, or 137 m$^2$ arcsec$^2$ per fiber.\footnote{This
assumes a telescope clear aperture with an effective diameter of 3.186
m based on a 3.499 m diameter primary and a central obstruction
diameter of 1.446m, as listed on
http://claret.kpno.noao.edu/wiyn/wiynfacts.html.}  The minimum fiber
spacing (center-to-center) is 5.63 arcsec due to buffer and
cladding. Figure 2 shows the astrometric format of the array. Active
fibers are sparsely placed in a nearly $2\times 2$ arcmin square array
of 367 mechanical-packing fibers. Seventy-five active fibers form a
$72.0^{\prime\prime} \times 71.3^{\prime\prime}$ sparsely-packed grid
offset into one corner of the array. This forms the ``object'' area of
the array. Except for a core region, one out of every three fibers in
this grid is an active fiber. The core region is densely packed with
17 fibers and has a fill-factor of 64.5\% within areas defined by 7
hexagonally packed, contiguous fibers (equivalent to a a filling
factor of $\sim 55$\% over an angular extent of $39^{\prime\prime}
\times 24^{\prime\prime}$).  The over-all fill-factor of the object
area is 25.4\%. The final seven active fibers serve to sample the
sky. They are well-spaced along two adjacent edges of the grid in an
L-shaped pattern, between 61.9$^{\prime\prime}$ and
85.6$^{\prime\prime}$ from the central fiber of the grid. There is a
$\sim$25$^{\prime\prime}$ gap between the edges of the grid and the
sky fibers. These ``sky'' fibers are uniformly distributed across the
fiber slit which feeds the spectrograph.

\subsection{Comparison to DensePak and Other Bi-Dimensional Spectroscopic
Instruments}

SparsePak complements the existing WIYN/DensePak array (Barden \etal\
1998). DensePak has 91 fibers, each 2.81$^{\prime\prime}$ in diameter,
arranged in a $7\times 13$ rectangular (but hexagonally packed)
array. DensePak fibers are spaced at 3.75$^{\prime\prime}$,
center-to-center, such that the array has the approximate dimensions
of $27^{\prime\prime} \times 42^{\prime\prime}$ on the sky at the
f/6.3 Nasmyth port. While SparsePak has coarser and sparser sampling
relative to DensePak, SparsePak has a factor of 3 larger grasp and
roughly 5 times the sampled angular foot-print.

SparsePak's enhanced grasp makes it well suited for studies of
extended sources at low surface-brightness, particularly because
comparable spectral resolutions are achievable with SparsePak as would
otherwise be obtained with DensePak. This is made possible by using
spectrograph settings where the anamorphic demagnification is large
(see \S2.2 below and Paper II). It is just these configurations that
yield the medium spectral resolutions (5000-25,000) that we desire for
galaxy kinematic studies. At these resolutions, sources with narrow
emission-lines can even be studied in bright time in the red, with a
minimum penalty payed for higher continuum background. However, the
key gain of SparsePak over DensePak is the ability to stay
background-limited at these medium spectral resolutions in reasonable
exposure-times.

Figure 1 illustrates the grasp versus the spectral power for a
representative sample of current, bi-dimensional spectroscopic
systems. The spectral power is the product of the spectral resolution
and the number of spectral resolution elements sampled by the
spectrograph. Spectrographic instruments on 8m-class telescopes
include Gemini's GMOS (Allington-Smith \etal\ 2002), VLT's VIMOS (Le
F\`{e}vre \etal\ 2003) and GIRAFFE/ARGUS (Pasquini, L. \etal\
2002). Spectrographic instruments on 4m-class telescopes include WHT's
SAURON (Bacon \etal\ 2001a) and INTEGRAL (Arribas \etal\ 1998), CFHT's
OASIS (Bacon \etal\ 2001b), Calar Alto's PMAS (Kelz \etal\ 2003), and
WIYN's DensePak and SparsePak. Interferometric instruments include the
Rutgers Fabry Perot Interferometer (RFPI), used on the CTIO 1.5m and
4m telescopes (Schommer \etal\ 1993, Weiner \etal\ 2001), and GHASP
(Garrido \etal\ 2002), used on the OHP 1.93m telescope. Figure 1 shows
the wide range in trade-offs made by instruments in terms of spatial
and spectral information collection, as discussed in \S1.

SparsePak falls in a unique location in the parameter space of Figure
1, having both one of the larger total grasps {\it and} spectral power
and the highest specific grasp (grasp per spatial resolution element)
of any instrument.  This superior performance comes at a cost, namely
in spatial resolution. In contrast, 4m-class instruments such as
SAURON and OASIS are optimized for higher angular resolution. This too
comes at a cost. These instruments sacrifice being able to achieve
higher spectral resolution for sky-limited observations. Even
instruments on 8m-class telescopes are unable to achieve the specific
grasp of SparsePak. This places SparsePak in a position to achieve the
highest possible spectral resolutions for sky-limited observations.

\subsection{The Bench Spectrograph}

The Bench Spectrograph is a bench-mounted, fiber-fed spectrograph
situated in a climate controlled room two stories below the telescope
observing floor. The spectrograph can be optimized for a wide range of
gratings because of its adjustable camera-collimator and grating
angles, and adjustable grating-camera distance.

The existing grating suite includes 6 low-order gratings with rulings
between 316 and 1200 and blazed between 4 and 31 degrees, and an R2
echelle (316 l/mm blazed at 63.4$^\circ$). These are used,
respectively at nominal camera-collimator angles of 30 and 11 degrees.
Although this angle is adjustable, in practice this option has not
been exercised. The echelle grating delivers between 2-5 times higher
spectral resolution than the low-order gratings, while the latter
deliver greater efficiency and increased spectral range at the cost of
resolution. Consequently the delivered product of spectral-resolution
$\times$ slit-width, R$\phi$, and throughput have a wide range of
values. This is quantified in Paper II.

There are two features of note concerning the Bench grating suite,
relevant to SparsePak's intended science mission. First, there are
large anamorphic factors for the echelle and low-order gratings blazed
above 25$^{\circ}$, such as the 860 l/mm grating.  With the echelle
grating the anamorphic demagnification is significant such that with
even the 500 $\mu$m SparsePak fibers the delivered instrumental
spectral resolution, $R$, is $\sim$10,000, with the FWHM sampled by
3.5 pixels. This is equivalent to a velocity resolution of 12.7 km
s$^{-1}$ ($\sigma$).

Second, the echelle grating is used in single-order mode, i.e., there
is no cross-dispersion. Orders are selected via rectangular narrow-band
interference filters placed directly in front of the fiber feed. These
filters have efficiencies of 90\% redwards of 600nm, 80\% around
500nm, and dropping to 60\% only as blue as 375nm. In the red these
values are considerably higher than what is achieved with reflection
gratings and hence for limited wavelength coverage in the red, the
Bench has the potential to out-perform grating-cross-dispersed
echelles.

$>>>>$

Despite this potential, the Bench Spectrograph total system throughput
(atmosphere, telescopes, fibers, spectrograph optics and CCD) is
estimated to {\it peak} at 5\% when using the multi-object fiber feeds
(Hydra Users
Manual\footnote{http://www.noao.edu/kpno/manuals/hydraman/hydrawiynmanual.html}
This value should apply to DensePak as well.  Measurements show
SparsePak's {\it peak} is 40\% higher, or roughly 7\% (see Paper
II). A significant portion of this gain is from decreased vignetting
in the fiber ``toes'' due to our redesign, as discussed in \S4.4. The
{\it mean throughput}, averaged over all fibers and wavelengths within
the field, is significantly lower, and closer to 4\% with SparsePak
and 2.5\% with the HYDRA and DensePak cables (see Paper II). The lower
mean values are due to the strong spatial and spectral vignetting,
which are severe for lack of proper pupil placement.  When using the
echelle grating, the large ($\sim$1m) distance between camera and
grating required to avoid vignetting the on-axis collimated beam
incident on the grating), the off-axis vignetting is particularly
large. In this mode, both spatial and spectral vignetting functions
contribute about a factor of 2 at the edge of the field such that the
slit ends at the end of the spectral range is down by typically
factors of 4 from the peak. We have taken this limitation into account
when mapping fibers from the telescope focal plane onto the slit (\S
3.2.6).

It's worth noting why the spectrograph has such severe vignetting. In
addition to the lack of pupil re-imaging, the spectrograph was
designed for a f/6.7 input beam and a 152mm collimated beam over a
modest field.  As we will show, the output from the fibers fed by the
telescope at f/6.3 are beams with focal ratios between 4.3 and 5.9 at
90\% encircled energy (EE). This results in collimated beams of 170 to
234 mm at 90\% EE.  While the optics are nearly sufficient for the
on-axis fiber\footnote{The effective clear apertures are 235 mm
diameter for the collimator; $203 \times 406$ mm for the echelle; $203
\times 230$ mm for the low-order gratings; and 206 mm diameter for the
camera. The projected grating areas for on-order settings with the
nominal camera-collimator angles are roughly $205 \times 210$ mm for
the low-order gratings blazed below 25$^{\circ}$, and only $205 \times
165$ mm for the echelle and low-order gratings blazed above
25$^{\circ}$.}, the slit sub-tends a 4.2 degree field on the parabolic
collimator; the fast output beams lead to loses in the system for the
off-axis fibers, compounded by the lack of a properly placed
pupil. Because the fiber-feed is in the beam, even the on-axis fibers
suffer some ($\sim$9\%) vignetting. As we show in Paper II, these
geometric considerations allow us to accurately model the observed
system vignetting. While the problem is currently severe, our clear
understanding of the problem gives good reason to believe that
significant improvements in the spectrograph optical system can be
made in the near future.

\subsection{Efficacy}

Given the Bench Spectrograph's low throughput, would a
higher-throughput, long-slit system be more competitive than
SparsePak? The long-slit, for example, would be stepped across a
source in repeated exposures. There are three primary reasons why
SparsePak will far outperform a long-slit spectrograph:

\begin{itemize}

\item The equivalent number of long-slits is $\sim$15, requiring a
throughput of the combined spectrograph, telescope and atmosphere of
105\%! This is a factor of 3 to 5 times higher than what can be
accomplished with the best, contemporary systems. While the improved
filling-factor of stepped, long-slit observations is equivalent to 3
SparsePak pointings (see \S 5), truly integral-field spectroscopy is
not needed for all applications. Hence only in the most extreme
scenario (integral-field spectroscopy is needed and a long-slit system
has 35\% efficiency), will long-slit observations break even in terms
of efficiency.

\item The long-slit observations would not be simultaneous and hence
conditions may vary, leading to uncertainties in creating
spectrophotometric maps.

\item The astrometric registration of stepped, long-slit observations
would be less certain.

\end{itemize}

Finally, as a general statement of cost-effectiveness, it should be
cheaper to build or upgrade a wide-field, high-resolution spectrograph
that is bench-mounted and fiber-fed rather than a direct imaging
system attached to a telescope port.

\section{DESIGN}

\subsection{Science Drivers}

Our aim is to provide a survey engine capable of measuring nearby
spiral-galaxy kinematics over most of the optical disk for the purpose
of determining their dynamics and their luminous (stellar) and dark
content. Since the distribution of mass can only be directly measured
by dynamical means, spatially-resolved galaxy kinematics provide
direct constraints on the origin and evolution of disk galaxies.  

In order to study the large-scale dynamics of the optical disks of
galaxies, the disks must be spatially well-sampled with spectroscopic
measurements out to several scale lengths, $R_s$. These measurements
must be at sufficient spectral resolution and signal-to-noise (S/N) to
determine both precise rotation, non-axisymmetric bulk motions, as
well as velocity dispersions in both gas and stars.  In addition to
spatial and spectral sampling and coverage, further technical
requirements include minimizing systematic errors due to cross-talk
and sky-subtraction. We discuss these science-driven technical
requirements in turn.

\subsubsection{Spatial Resolution and Sampling}

Consider first the spatial resolution and sampling requirements.
Typical disks have exponential scale-lengths of 2-5 kpc in size. To be
well sampled, there should be 2-3 measurements per disk scale-length.
In the inner regions, where the rotation curve rises and changes shape
rapidly, the sampling should be finer by additional factors of 2-3.
In the outer regions, where the disk is fainter, it is important to
have more solid angle sampled such that the limitations of decreased
surface-brightness can be overcome by co-adding signal, e.g., within
annular bins. A generic, scale-free requirement, therefore, is for
enhanced resolution at small radii, and enhanced coverage
(solid-angle) at large radii. The latter is naturally achieved by a
two-dimensional sampling pattern.

The absolute spatial scale is set by disk-galaxy structure, which, if
not fully understood from a theoretical perspective, is at least
observationally well defined. Two to three disk scale-lengths
represents a threshold for the mass distribution within spiral
galaxies in terms of transitions between different components of the
overall mass distribution. At these distances rotation curves are
expected to be flat, or at least have transitioned from the steep,
inner rotation-curve rise, to a more shallow rise or fall. Hence, with
rotation curves extending out to these radii, one may suitably
estimate a terminal rotation velocity and total dynamical mass. Since
the disk is expected to contribute maximally to the overall enclosed
mass budget near 2.2 $R_s$ (Sackett 1997), dynamical disk-mass
estimates need to probe out to at least these distances.

For a finite number of fibers of a {\it fixed} physical size on a {\it
given} telescope, the above requirements imply a sampling area,
resolution and pattern that is highly specific for galaxies of a
particular angular size. The only way to substantially increase the
dynamic range in this case is to modulate the input plate scale via
fore-optics (which may include lenslets). Such optics introduce
additional light-losses both from reflections and, in the case of
lenslets, from mis-alignment or non-telecentricity (Ren \&
Allington-Smith 2002). Plate-scale modulation is limited by the
numerical aperture of the fibers (roughly f/1.3 to f/2 at the coarse
limit), and by the need to feed the fibers at sufficiently fast
$f$-ratio to avoid introducing significant FRD (roughly f/4 to f/6 at
the fine limit). At fine plate scales the grasp is decreased, and
consequently so too is the achievable depth. Alternatively, for a
given spectrograph, one may choose the largest possible fibers that
maximize the grasp in the absence of fore-optics, while yielding the
required spectral resolution. Targets can then be chosen to suit the
above sampling criteria. Since galaxies are found in a wide range of
apparent sizes, we chose this latter path.

For a fixed-scale integral field unit (IFU), it is still possible to
fine-tune the spatial sampling geometry to allow for some dynamic
range in spatial scale.  Indeed, there is recourse in carefully
designing a sampling pattern to be coupled with specific observational
techniques (i.e., dithering patterns). Herein lies a critical
advantage of fibers for formatted patterns, or what we call
``formatted field units'' (FFU).  In other words, since fibers are
convenient light-pipes, it is not necessary to sample truly-integral
regions of the sky, but instead one can consider optimal geometries to
accomplish a specific science goal. 

Our original pattern consisted of four, rotated long-slits (at
position angles of 0$^{\circ}$, $\pm 30^{\circ}$, and 90$^{\circ}$)
and an integral, inner region (Bershady 1997, Bershady \etal\ 1998,
Andersen \& Bershady 1999).  However, these designs required rotation
to fill in interstitial regions, and did not provide more sampling of
solid angle at larger radii.  Ultimately the FFU concept led us to
develop a pattern with wide areal coverage with sparse sampling in a
rectangular grid, combined with a densely sampled core, as per the
above desiderata.  The pattern described in \S 2 allows for simple
dithering to either fill the sparsely sampled grid or to critically
sub-sample the core, as discussed in \S5. The rectangular grid is also
convenient for tiling of very large sources. Further, a rectilinear
sampling provides a variety of radial samplings when centered on an
axisymmetric source.  The final pattern is well suited for study of
normal, luminous spiral galaxies with recession velocities of
(roughly) 2,000-10,000 km s$^{-1}$.

\subsubsection{Spectral Resolution}

The second consideration is the spectral resolution required to
measure disk kinematics. While disks have typical rotation velocities
of order 100 of km s$^{-1}$, the non-axisymmetric motions are of order
10 km s$^{-1}$, as too are the velocity dispersions in both gas and
stars. In particular, the vertical component of the stellar velocity
ellipsoid, $\sigma_z$, is expected to be of order $\sim$10 km s$^{-1}$
for the outer parts of disks, based on what we know of disk stars in
the Solar Neighborhood, and from long-slit measurements of a handful
of nearby galaxies (Bottema 1997). In order to optimize the
measurement of $\sigma_z$ in galaxies of known rotation velocity,
nearly face-on galaxies must be targeted.  This optimizes the
projection of $\sigma_z$, but minimizes the projection of the rotation
velocity. This is a reasonable trade since the rotation velocity is
typically an order of magnitude larger than the velocity dispersion.

While it is possible to centroid a high S/N line to better than 10
times the instrumental resolution, the same precision cannot be
achieved (at a given S/N) for the higher-order moments of line-width
($\sigma$) skew, and kurtosis (the latter two are equivalent to the
Gauss-Hermite-polynomial h$_3$ and h$_4$ terms).  Intuitively, one may
understand that higher-order profile information requires better
resolution or better S/N.  Consequently, the desired
precision of the velocity dispersion measurements provides the driver
for the required spectral resolution.

To reliably measure velocity dispersions of 10 km s$^{-1}$ we estimate
that instrumental resolutions of $\sim$10,000 are necessary. This
statement is qualified by the obtainable S/N. A full treatment of the
trade-offs of profile-moment precision versus S/N is beyond the scope
of this work. However, we find that S/N of 15 to 20 in a spectral line
yields line-width measurements at a precision of 10\% for widths at the
instrumental resolution. Given the trade-offs between grasp and
spectral resolution (larger fibers collect more light but yield lower
spectral resolution) we estimate that absorption-line S/N of greater
than 20-30 is unlikely to be obtained in the outer parts of disks for
any reasonable exposure times on 4m-class telescopes.  Consequently it
is not possible to push too far below the instrumental resolution
for any reasonable precision. Hence for stellar velocity dispersions
studies in dynamically cold disks, adequate spectral resolution is at
a comparable premium to S/N and spatial resolution.

\subsection{Practical Constraints}

The design of SparsePak is constrained by mating to an existing
telescope feed and spectrograph. For practical purposes we accepted
the envelopes imposed by this existing hardware. Our adopted
fabrication process also imposed certain practicalities. We mention
here those constraints which are relevant to placing limits on our
science goals.

\subsubsection{Fiber Size}

Based on our experience with fibers, we find that 500 $\mu$m is a
maximum practical thickness of the active diameter in order that the
fiber stiffness does not cause frequent breakage in
handing. Fortuitously, this corresponds to the maximum size we would
want to consider based on our interest to achieve spectral resolutions
of order 10,000 using the echelle grating with the Bench Spectrograph.

\subsubsection{Spectral Coverage}

Because SparsePak is built for an existing spectrograph, spectral
coverage is not a design issue per se. However, we did consider
whether the Bench Spectrograph's spectral coverage was suitable for
our science goals. Given the large suite of gratings, a wide variety
of spectral coverage is available. For kinematic studies, however, we
are interested in the higher-dispersion gratings.  What is relevant,
really, is the number of independent resolution elements,
$N_{\Delta\lambda}$, which, depending on the setup (i.e., the degree
of demagnification) is between 600 and 800. In general for spectral
resolutions, $R$, the spectrograph will cover a spectral range of
$N_{\Delta\lambda} \lambda / R$. For $R = $ 10,000, the covered range
is several hundred Angstroms in the optical.  This is sufficient to
cover the Mg~I region from [O~III]~$\lambda$5007\AA\ past the Mg~I$b$
triplet and [N~I] $\lambda$5200 stellar absorption lines; or the
H$\alpha$ region from [N~II]~$\lambda$6548 to [S~II]~$\lambda$6731; or
all three lines of the Ca II near-infrared triplet at 8498, 8542, and
8662 \AA.

Greater spectral coverage is generally advantageous for
cross-correlation work in weak-line regions, such as the Mg~I region
near 5130\AA, since the desired power in the cross-correlation comes
from many weak (e.g., Fe~I) lines spanning a wide range of
wavelength. For a finite detector focal-plane, increased spectral
coverage comes at the cost of decreased spectral resolution or spatial
coverage. The optimum trade is highly dependent on the scientific
goals, but it is unlikely that the current system is far off for
studies of stellar kinematics in galaxies.

\subsubsection{Cross-Talk}

Integral-field spectroscopy is likely to have cross-talk between
individual spatial channels in the telescope focal plane due to the
blurring effects of the atmospheric point-spread-function.  While
there is no indication of fiber-to-fiber cross-talk for the types of
fibers we have used (i.e., photons do not leak out of the fiber
cladding and penetrate the cladding of a neighboring fiber), for
fiber-based integral-field units, there is an added consideration.

Because of the azimuthal scrambling in fibers, and the requisite
remapping of the two-dimensional telescope focal-plane into the
one-dimensional spectrograph slit, nonadjacent fibers in the telescope
focal-plane will be adjacent in the slit. This can lead to spatially
incoherent, but systematic cross-talk. To minimize systematic effects
it is therefore desirable to adequately separate fibers along the
slit. The specific fiber separation depends on the optical quality of
the spectrograph optics (both aberrations and scattering), as well
as the scientific need to control the level of systematics.

For SparsePak, given the large fiber buffers (0.9 arcsec edge-to-edge
for the most closely packed fibers) and excellent WIYN image quality
(a median seeing of 0.8 arcsec FWHM), there is very little coupling
between fibers in the telescope focal plane. Hence the only
significant cross-talk would take place at the spectrograph slit.
Because of the difficulty in assessing the effects of systematic
errors due to cross talk, we have chosen conservative limits. Our
adopted science requirement is to limit cross-talk to $<$1\% for
discrete spectral features from adjacent fiber channels. This limits
systematic effects to 10\% for adjacent fibers with factors of 10
difference in signal flux. Such variations in signal are likely
worse-case given the fiber mapping and the astrophysical variations
of, e.g., H$\alpha$ emission within galaxy disks.  Incoherent
cross-talk (i.e., scattering into and out of the source spectrum)
limits are $<$10\%, with a goal of $<$1\%. This component mainly
affects the delivered S/N in an r.m.s. sense. Because incoherent
scattering takes place over larger physical scales, it is dominated by
the spectrograph optics.  Fiber separation (\S 3.2.4) is designed,
then, to meet the requirement for coherent scattering even when
on-chip binning by factor of 2 in the spatial dimension. (On-chip
binning is important for low light-level applications, and provides
significant gains given the large projected fiber diameter onto the
CCD -- roughly 4 unbinned pixels in the spatial dimension.)

\subsubsection{Total Fiber Number}

At the spectrograph input focal plane, the maximum slit-length
currently used by any of the fiber feeds\footnote{These include the
red and blue Hydra MOS cables, DensePak, and now SparsePak.} is
76.4mm.
There is also a minimum fiber-to-fiber separation at the spectrograph
feed to prevent significant cross-talk between fibers. This is a
function of the scattering properties and image quality of the
spectrograph. To meet the design requirements of \S 3.2.3, we
estimated $\sim 400 \mu$m was the minimum acceptable edge-to-edge
distance between the active regions of fibers at the spectrograph
slit. This estimation was based on the performance of 3 existing fiber
feeds for the WIYN Bench Spectrograph.  A detailed measurement of the
SparsePak cross-talk is presented in \S 5 of Paper II.

The above combination of maximum slit-length, minimum fiber separation
at the slit, and maximum fiber size constrains the total number of
fibers and hence the overall maximum grasp of the system. A maximum of
82 fibers with 500 $\mu$m diameter cores was chosen. These fibers map
into a 73.6mm slit. An additional 2-3 fibers could be added to bring
the slit-length up to the nominal 76.4mm value of the other Bench
cables. However, due to the strong vignetting within the spectrograph,
the addition of extra fibers offered little gain.

\subsubsection{Array Size}

At the telescope focal-plane we are limited by the existing telescope
mounting hardware. The entire fiber array assembly (array plus its
mount) must fit within a cylindrical mount with a one-inch (25.4mm)
outer diameter.  The array had to be rectangular in cross-section
given the way in which it was glued (as described in \S4.3.2).  This
yields a maximum array dimension (cross-section) not to exceed
$\sim$12mm, which corresponds to a maximum field of view of 112 arcsec
(diameter), with diagonals up to 160 arcsec.

This limiting field-of-view of the FFU is comparable to the size of
nearby normal, luminous spiral galaxies. To maximize the distance
between the object grid and the sky fibers, the object grid is placed
in one corner of the fiber array, and the sky fibers are then placed
in an L-shaped pattern around the two far sides (see Figure 2).

\subsubsection{Minimizing the Effects of Vignetting}

The mapping of fibers between telescope and spectrograph input focal
planes is complicated by several redesigns of the active fiber lay-out
during construction. However, one of the goals was to put some of the
fibers in the center of the source grid near the outside of the
spectrograph slit, and vice-versa, the reason being that since
astronomical sources are generally centrally concentrated, this would
balance the strong vignetting in the spectrograph.
Ideally, we would have adopted a more ordered mapping (e.g., Garcia
\etal\ 1994), but the large fiber diameter and fiber-to-fiber
separation makes the details of the mapping largely unimportant.

\subsection{Sky Subtraction}

Random and systematic errors in sky-subtraction have plagued fiber-fed
spectroscopic measurements. Here we motivate our fiber allocation
calculated to minimize random errors, and discuss how careful
placement and treatment of sky fibers in the spectrograph and
telescope focal planes help limit systematic errors.

\subsubsection{Optimum Number of Sky Fibers}

Wyse \& Gilmore (1992) calculate the optimum allocation of fibers to
source and sky for the particular case of random errors where source
flux and sky flux are equal. Here we consider a similar calculation
but for the two extreme cases of background-limited and
detector-limited observations. These are more relevant for
observations at low surface-brightness and high spectral
resolution.\footnote{Not considered here is an additional source of
error, relevant to high-signal regimes, from imperfect pixel-by-pixel
correction of the CCD response, i.e., field-flattening.}

The adopted merit function assumes one is trying to achieve a
specified S/N for a given number of sources ($N_{source}$) in the
least amount of total observing time ($t_{total}$). (Here, S/N can be
defined as any linear function of the S/N per recorded detector
element.) Observation of these $N_{source}$ sources constitute a
``survey.'' In other words, one would like to maximize the merit
function $f_{merit} = N_{source} / t_{total}$. Further, we assume that
a spectrograph is fed by a finite number of fibers, $n_f$, that can be
used for any given observation; that some number ($n_s$) of these
fibers will be used for sky; that a survey may consist of more than
one observing set (e.g., $N_{total} > n_f - n_s$); and that sky can be
subtracted perfectly -- in a statistical sense, i.e., sky contributes
to shot-noise, but not to systematic error.

In the background- and detector-limited regimes, 

\begin{equation}
S/N \ \propto \ \left\{ \begin{array}{ll}
        \sqrt{\frac{t}{1 + 1/n_s}} & \mbox{background limited} \\
        \frac{t}{\sqrt{1 + 1/n_s}} & \mbox{detector limited} \\
        \end{array} \right.
\end{equation}

\noindent where $t$ is the observing time for a given source, which
can be expressed as

\begin{equation}
t \ = \ \frac{t_{total}}{N_{source} / (n_f - n_s )}.
\end{equation}

\noindent These equations can be combined to solve for the survey
merit function:

\begin{equation}
f_{merit} = \frac{N_{source}}{t_{total}} \ \propto \ \left\{ \begin{array}{ll}
        \frac{n_f-n_s}{1+1/n_s} & \mbox{background limited} \\
        \frac{n_f-n_s}{\sqrt{1+1/n_s}} & \mbox{detector limited} \\
        \end{array} \right. 
\end{equation}

\noindent Maximizing the merit function with respect to $n_s$ (at fixed
$S/N$ and $n_f$) yields quadratic relations
with these exact solutions:

\begin{equation}
n_s \ = \left\{ \begin{array}{ll}
         -1 + \sqrt{1+n_f} &  \mbox{background limited} \\
         \frac{3}{4} ( \ -1 + \sqrt{1 + \frac{8}{9} n_f} \ ) & \mbox{detector limited} \\
        \end{array} \right.
\end{equation}

\noindent which are plotted in Figure 3.

Equation (4) is a general result for background- and detector-limited
surveys, which are essentially identical.  This result is independent
of spectral resolution, and independent of whether source fibers
target many, individual targets, or are bundled into a single IFU
targeting one, extended source. Examination of Figure 3 indicates that
SparsePak, with 82 total fibers, should have of order 8 sky
fibers. For reasons of symmetry in the object grid, we chose to
allocate 7. In contrast, DensePak has 4 allocated sky fibers, whereas
the optimum number is closer to 9. So while it may appear that
DensePak is more efficient by allocating fewer fibers to sky, from a
survey perspective SparsePak is closer to the ideal. However, the
merit function is not strongly dependent on the number of sky
fibers. The value of the merit function for DensePak, for example, is
only 6\% lower than its optimum value.

%

\subsubsection{Slit-Mapping}

Our argument in the previous section does not take into account
instrumental issues which affect the final data quality and ability to
extract signal accurately. While scattered light plays an important
role in the ability to accurately subtract background continuum, the
primary contribution to systematic errors in the subtraction of
spectrally unresolved sky lines are the field-dependent optical
aberrations present in spectrographs (Barden \etal\ 1993b). The WIYN
Bench Spectrograph, for instance, uses a parabolic collimator with
field angles ranging from 0 to 2.1$^\circ$. Evidence for
field-dependent effects are shown by Barden \etal\ (1993a) for the
Mayall 4m RC spectrograph, and also are evident in the 2dF system as
reported by Watson \etal\ (1998): With sky fibers concentrated in one
area of of the slit, sky residuals increase for fibers farther away
along the slit.

One technique to deal with the issue of aberrations is known as
``nod-and-shuffle'' (Glazebrook \& Bland-Hawthorn, 2001), whereby the
telescope is nodded between source(s) and sky at the same time that
the charge is shuffled accordingly on the detector.  While initially
presented in the context of multi-slit spectroscopy, nod-and-shuffle
can be applied to multi-fiber spectroscopy as well.  This technique
has the advantage of putting both sky and source flux down the same
optical path, while sampling both over the same period of time.  The
more traditional ``beam-switching'' technique (e.g., Barden \etal
1993b), for example, suffers from the inability to sample source and
sky at the same time. However, both techniques suffer from allocating
50\% of the observing time to sky. This is equivalent to allocating
half of the total fibers to sky. While nod-and-shuffle undoubtedly
achieves the smallest level of systematic error, the penalty in terms
of the above survey merit function (random error) may be too high. An
alternative approach may be to try and map the optical aberrations
within the spectrograph system (e.g., Viton \& Milliard, 2003).

For our purposes, implementing nod-and-shuffle is beyond the scope of
the current effort.  We have chosen instead to carefully place our 7
sky fibers such that they sample at nearly uniform intervals along the
slit. With the reasonable assumption that the optical aberrations are
symmetric about the optical axis, we expect to be able to model the
aberrations empirically via these small number of sky fibers. The
success of this method, and its dependence on any differential FRD
within the fibers, is demonstrated in Paper II. At present what is
relevant for the instrument design is the concept of mapping the sky
fibers in the focal plane across the slit.

\subsubsection{Sky-Fiber Placement Within the Fiber Array}

Our experience with DensePak (Andersen 2001) indicates DensePak sky
fibers behaved differently then those sampling the source. In this
FFU, the ``source'' fibers are glued together coherently into a
rectangular array, while the sky fibers are separately mounted in
hypodermic needles, and offset from the array. The differences we
found were such that the continuum levels measured in the sky fibers
were systematically above or below the continuum in the ``source''
fibers after field-flattening, even when the array was pointed at
blank sky. While we never determined the exact cause for this
systematic behavior, it seemed reasonable to suppose that differences
in fiber termination may have played a role.  For this reason, we
designed SparsePak to include the sky fibers within the same coherent
fiber bundle as the source fibers.

\subsection{Summary of Design Considerations}

The final SparsePak design was dictated by a confluence of, and
compromise between scientific objectives, technical performance goals,
and mechanical and fabrication constraints. Within the confines of the
existing spectrograph and telescopes feed, the spatial and spectral
sampling are the key drivers which determined the fiber size and
lay-out of the SparsePak formatted field unit. Cross-talk was a
secondary condition which provides some limits on the fiber packing
and hence total number of fibers. Sky subtraction dictated some
additional fine-tuning of the fiber allocation and placement. The
above discussion provides generic requirements to yield adequate
observational data for a wide range of dynamical studies in the
context of the practical constraints of the WIYN telescope feed and
existing Bench Spectrograph.

\section{TECHNICAL SYNOPSIS}

Summarized here are the technical attributes of the SparsePak cable
detailed in the Appendix deemed directly relevant for its performance.

\subsection{Head Construction: Buffering}

The fiber head has short, ``packing'' fibers surrounding the long,
active fibers, cut from the same Polymicro batch. These serve as
mechanical elements, and provide an edge buffer with a minimum
thickness of one fiber. The buffer is intended to minimize stress on
the active fibers, and maximize their condition uniformity. The
success of this buffer arrangement is evaluated in \S 6.

\subsection{Cable Design}

The cable consists of an outer sheath of heavy-gauge flexible
stainless-steel conduit and an inner PVC tube jointed every 6 feet to
provide natural spacing within the larger stainless conduit. Within
the PVC cable run 82 black Teflon tubes (each containing 1 fiber). The
stainless-steel flex-conduit serves to protect against fiber crushing
and over-bending. The PVC and Teflon provide safe, smooth inner
surfaces for the Teflon and fiber, respectively. We believe this
design is successful in minimizing stress-induced FRD {\it along} the
cable length, although the addition of thermal breaks in the Teflon would
be advantageous in future designs (Fabricant \etal\ 1998).

\subsection{Cable Termination and Interfaces}

The cable is terminated with mounts whose design are dictated to a
large extent by existing mounting hardware in the telescope and
spectrograph focal surfaces. Three significant modifications were made
within these constraints. (1) To ensure and maintain telecentricity of
fibers in the telescope focal plane the head-mount dimensions were
precisely machined, and a support brace attaching to WIFOE was made.
(2) An anti-rotation collar is placed roughly 250 mm back from the end
of the fiber head to prevent the bare fibers from twisting. (3) The
mount to the spectrograph has a modified slit-block, and the exit
aperture of the filter-holding ``toes'' has been enlarged
considerably. The latter allows up to an f/4 unvignetted beam to exit
the fibers into the spectrograph. Measurements presented in Paper II
show that this enlargement may increase the throughput by $\sim$20\%.

One last feature of the existing mounting hardware to note here is the
fiber foot (where the cable terminates for mounting on the
spectrograph). As we evaluate in \S 6, this curvature is too sharp,
and is the principal cause of FRD in the system.

\subsection{Reference Cables}

To determine the effects of the cable manufacturing process specific
to the FFUs on fiber throughput and FRD, and to provide a stable
reference for future testing, we produced several, single-fiber
``reference'' cables. Two of these cables were made from the 500$\mu$m
fiber -- a ``short'' cable, 1.5m length, and a ``long'' cable, 24.5m
in length. The last meter of each cable is covered with the identical
black, opaque Teflon used in the FFU cables. The remainder of the
fiber is uniformly coiled on the initial foam packing-spools on which
the fiber came. The fiber ends are all terminated inside a micro-tubes
of appropriate diameter, and glued with a single drop of Norland 68 UV
curing epoxy. These tubes are themselves glued into machined, brass
ferrules suitable for mounting on an optical bench with standard
hardware, or into our circular-lap polisher. The fiber polishing
process is identical to that used for the FFU cables on this
polisher. As expected, no hand-polishing was necessary.

These reference cables represent idealized application of astronomical
fiber light-conduits in that they have excellent polish, the glue type
is superior, the glued surface-area is minimal, and there is otherwise
little stress (or change in stress) on the optical fibers.

\section{ASTROMETRIC SPECIFICATIONS}

Final SparsePak head dimensions were measured in our lab independently
by two, skilled technicians, each using two different micrometer
engines. The final SparsePak array of $23 \times 20$ fibers is very
close to 12.05mm square at the front face. This maps to
113.0$^{\prime\prime}$ at WIYN IAS port assuming the nominal plate
scale of 9.374 arcsec/mm. The precise dimensions are $12.09 \pm
0.05$mm in width, and $12.01 \pm 0.03$mm in height, as detailed in
Figure 2.\footnote{The width dimension is defined as the 23 rows in
direction orthogonal to major axis of central, densely packed
fibers. Mechanically, the width is in the direction of the cut in the
head-mount that forms the clamp. The height dimension is defined as
the 20 rows in direction parallel to major axis of central, densely
packed fibers. Mechanically, the width is in the direction orthogonal
to the cut in the head-mount that forms the clamp.} The array is
square to within $0.6 \pm 0.3$\%. The array dimensions imply average
fiber-to-fiber separation at the face of 525.6 and 600.5 $\mu$m, or
4.927$^{\prime\prime}$ and 5.629$^{\prime\prime}$, respectively, in
width and height. The array dimensions also imply an average glue
thickness of 0.5$\mu$m where the fiber buffers abut. Measurements of
the array dimensions along the 50.8mm length of the glued volume
indicates flaring of $-0.13 \pm 0.01^\circ$ in width and $+0.10 \pm
0.07^\circ$ in height, where the sign of the flaring indicates whether
the flaring is toward ($+$) or away ($-$) from the central axis of the
fiber head. The amplitude of this bundle flaring is well under our
tolerance limit in terms of the FRD error budget: Differential
effects, center-to-edge, are well under a 0.1$^\circ$.

Astrometry based on direct imaging of the fiber face (e.g., Figure 11)
indicates that the fiber-to-fiber spacing is uniform within our
measurement errors (1\% of fiber width, or $<0.05^{\prime\prime}$).  A
table of astrometric positions of the fibers relative to the central
fiber (\#52), useful for creating maps of extended sources, is
available at the SparsePak web
site.\footnote{www.astro.wisc.edu/$\sim$mab/research/sparsepak} Two
common observing offset-patterns are also provided there.  The ``Array
fill'' pattern of three positions provides complete sampling at every
fiber position (e.g., every 5.6$^{\prime\prime}$) within the nominally
sparsely sampled 72$^{\prime\prime} \times 71^{\prime\prime}$
grid. This pattern is useful, for example, for creating velocity
fields of spiral disks (e.g., Andersen \& Bershady 2003, Courteau et
al. 2003, Swaters \etal\ 2003, Verheihen \etal\ 2004). The ``Array sub-sample'' provides
critical sampling, i.e., at every half-fiber position in both
dimensions (roughly 2.8$^{\prime\prime}$ spacing) within the
densely-sampled core. This pattern is useful to obtain the highest
spatial resolutions within the inner $39^{\prime\prime} \times
25^{\prime\prime}$ region of an extended object.  By combining
these two patterns (9 positions total), critical sampling is achieved
over the full $72^{\prime\prime} \times 71^{\prime\prime}$ grid.

\section{OPTICAL PERFORMANCE}

Prior to shipping and installing the SparsePak cable onto WIYN, we
characterized the completed cable on an optical test-bench in
our lab. The test-bench system was designed to measure absolute
throughput and FRD at a number of optical wavelengths for which we had
available filters.

\subsection{Optical Test-Bench}

The test-bench setup, illustrated schematically in Figure 4, consists
of a double re-imaging system using commercially available, 2-element,
50mm achromats. The concept is based on earlier systems developed by
Barden \& Ramsey, as reported by Ramsey (1988): A differential flux
and flux-profile comparator is made from two optical re-imaging
systems with an intermediate focus that can be switched between (1) a
``straight-through'' mode where the first re-imaging system directly
feeds the second, and (2) a ``fiber'' mode where the first re-imaging
system feeds a fiber which then feeds the second re-imaging
system. Modes (1) and (2) differ only by the presence of the optical
fiber inserted at the intermediate focus, which forms an optical
diversion adding zero net length to the imaging portion of the
system. The modes are selected by the simple translation of a
precision stage which holds the entire first re-imaging system and the
output end of the fiber. 

The first re-imager serves to place an image of a uniformly
illuminated pin-hole at an intermediate focus with a beam of known and
modulatable $f$-ratio (the ``input $f$-ratio''). As noted above, this
focus can be transferred either directly to the second re-imaging
system (``beam mode''), or into a fiber (``fiber mode''). In fiber
mode, the fiber feeds the second re-imaging system. In both cases, the
second re-imaging system transfers the intermediate focus to the
surface of a CompuScope CCD.\footnote{Precision Instrumentation \&
Software, Santa Barbara, CA.} This detector has a $768 \times 512$
format of 9$\mu$m square pixels. The second re-imaging system has a
known and modulatable ``output $f$-ratio.''

For both re-imaging systems, the $f$-ratio modulation is accomplished via a
graded iris placed in their respective collimated beams. Ideally the
iris would be placed at the pupil formed by the collimator lens, but
space limitations on our optical bench prevented us from doing
this. Given the small field used in the system, i.e., the image is a
pin-hole, the vignetting produced by our setup is negligible.  Because
the camera lenses are over-sized given the effective beam stops of the
irises, it is unimportant for the camera optics to be at the collimator
pupil.  Pellicles were inserted into the collimated beams of both
re-imagers for initial optical alignment (by visual inspection via a
telescope and by tracing via a laser feed). One pellicle was used
during the measurement stage in the first re-imaging system for
alignment of the intermediate focus onto the fiber.

Pin-holes were illuminated by a lamp via a coherent fiber bundle
illuminating a baffled diffuser, and then a filter, in that order. We
found this specific setup and careful baffling of the pin-hole
illumination was essential to minimize scattered light.  Due to their
small size, filters were placed between the baffled diffuser and the
pinhole.  Neutral density (ND) filters were also required since high
lamp intensities were needed for source-stability, optical alignment,
and to place the pin-hole image on the fiber face. Placement of the
NDs in front of the CCD considerably eased the measurement process.

The size of the pin-hole is dictated by the magnification of the first
re-imaging system and the desire to under- or over-fill the fiber
face. For the test-bench measurements reported here, we used lenses
with 250, 200, 150, and 100mm focal lengths at L1 through L4
respectively.  We chose a 400$\mu$ pin-hole for SparsePak such that
the re-imaged size at the intermediate focus was 320$\mu$m. As such,
this permitted us to illuminate a large fraction of the fiber face
while being sure that all of the incident flux went into the fiber. We
also tried a smaller, 10$\mu$ pin-hole to verify (a) that all of the
light was being fed into the fiber, and (b) that FRD measurements did
not depend on the specific input modes filled at constant $f$-ratio.
For example, with the smaller pinhole we were able to align the spot
on the middle and edge of the incident fiber face.  The results of
these tests with the smaller pinhole were positive, and so we focus
below on results using the 400$\mu$ pin-hole.

The filters available at the time of SparsePak testing consisted of a
``standard'' $UBVRI$ set. Narrower bandwidths are desirable,
particularly in the blue where fiber and CCD response change rapidly
with wavelength.

\subsubsection{Comparison to other FRD Measuring Engines}

The difference between our experimental design and earlier ones (e.g.,
Ramsey 1988) is in the details of the optical arrangement,
opto-mechanics of the alignment process, and the use here of an areal
detector instead of an aperture photometer. The use of a CCD reduces
sensitivity to defocus, permits a better understanding of optical
alignment and focus, and yields more accurate and precise estimations
of the total transmitted light (via the ability to perform
multi-aperture photometry and determine background levels).

However, we have not taken full advantage of the areal detector,
namely to image the far-field output pattern of the direct and
fiber-fed beams. This would directly allow us to measure the effects
of FRD on the beam profile in one step, i.e., we wouldn't need
measurements at multiple $f$-stops, as described in the next section.
Carrasco \& Parry (1994) have implemented such a scheme, effectively
by placing the CCD camera directly behind what would be our first
focus. The disadvantage of their scheme is that a precisely repeatable
back focal-distance must be ensured since they are imaging an
expanding beam.  

A viable alternative for future consideration is to place the CCD at
the pupil of what is our second collimated beam. In practice this
requires the necessary optics to make a small enough collimated beam
to match the available CCD, and possibly the addition of a field-lens
near the first focus (our Focus 1) to place the pupil at a
back-distance convenient for CCD mounting.

Such a system as we have just described may be competitive with, and
certainly complementary to the ``collimated beam'' approach described
by Carrasco \& Parry (1994). The latter uses a laser to directly probe
the FRD at a given input incidence angle, and then relies on a model
to synthesize the full effects of FRD on a astronomical beam profile
(i.e., a filled cone with obstructions). The approach described here
is model-independent, provides a means also for measuring total
throughput, and may provide a simpler and more cost-effective way to
measure the wavelength dependence of throughput and FRD.

\subsection{Laboratory Measurements}

For each filter and fiber, the idealized measurement process consisted
of (i) establishing the optical alignment of the system at a given
input $f$-ratio; (ii) obtaining a measurement of the beam-mode flux at
an output $f$-ratio of f/3 (``open''); (iii) transferring the setup to
fiber mode and carefully aligning the pin-hole image with the fiber
center; (iv) obtaining a series of fiber-mode flux measurements while
varying the output $f$-ratio from f/3 to f/12 (typically 5-10
exposures at a given $f$-ratio); (v) re-acquiring beam-mode and taking
an identical series of flux measurements. Post-acquisition
image-processing was done via IRAF, the goal of which was simply to
measure a total flux from the pin-hole or fiber-output image incident
on the CCD.

Considerable care was taken with mounting SparsePak and the reference
cable to ensure that the fibers were aligned to the optical axis
within 0.2$^\circ$. As with telescope alignment, even small off-axis
angles induce appreciable FRD. In the case of SparsePak, the mounting
hardware was considerable given the bulk and stiffness of the cable
and the need to actuate the slit between fiber and beam modes.

Due to the short time period between completing SparsePak's
manufacture, the final alignment of the test-bench, and the shipping
date, characterization of the SparsePak fibers were done within the
short period of 6 days between April 19 and April 24, 2001. Some of
these measurements are known to have been problematic in terms of
optical alignment. We were careful to note when we thought the
placement of the pin-hole image on the fiber was poor or uncertain, or
if other aspects of the optical set-up were questionable. With the
exception of lamp variability (which might produce errors of either
sign), all of the other systematics in our measurement process would
lead to {\it underestimating} the true throughput of the fibers. As we
will show, the measurements we were able to obtain give consistent and
plausible results that SparsePak is a high-throughput fiber cable with
explainable trends in FRD.

\subsection{Fiber Transmission}

We have measured the total fiber transmission for 13 SparsePak fibers
in the B, V, R, and I bands for an input $f$-ratio of 6.3. We have
also made identical measurements for the SparsePak reference fiber,
both at f/6.3 and f/13.5 input focal ratios. The SparsePak fibers were
chosen to lie over a range of positions within the SparsePak head as
well as to span the slit.  The total fiber transmission is defined to
be the light transmitted within an output beam of f/3. As we show in
the next section, the encircled energy as a function of output
$f$-ratio converges by this value.

Recall that fiber throughput measurements are done in a differential
way by comparing the total counts measured with the test-bench CCD in
``imaging'' and ``fiber'' modes. What we report, then, is a
measurement of the total fiber transmission which includes
end-losses. As noted, measurement systematics that we could identify
included lamp drift or poor alignment of the illuminated, re-imaged
pin-hole onto the fiber. For the latter we had to rely on detailed
measurement-log notes. For the former, we could check the stability of
the observed flux over a series of measurements at a fixed iris
aperture, as well as comparing initial and final beam-mode fluxes. Of
the 13 fibers measured, 5 fibers were flagged as being problematic:
\#37, 39, 72, 81, 82.

The results of our measurements are presented in Figure 5. The 8
fibers for which we have robust measurements in all bands are in
agreement with expected values based on the manufacturer's attenuation
specifications plus two air-silica interfaces -- at least at
wavelengths corresponding to $V$, $R$ and $I$ bands. In the $B$ band
our measurements appear too high. This is because we have not
correctly assessed the color terms of the band-pass, and hence the
proper effective wavelength for the measurements. We estimate that the
combination of the relatively cool lamp filament (due to modest lamp
intensity to optimize stability), red fiber transmission, and dropping
quantum efficiency of the detector in the blue yields an effective
wavelength closer to 4900\AA\ for the $B$ filter. From inspection of
Figure 5, one can see our measurements through the $B$ filter are in
agreement with the predicted performance assuming such a red effective
wavelength.

Finally, we found that the reference cable had systematically higher
transmission than the median value for the SparsePak fibers -- roughly
3-4\%. One might be tempted to conclude that this is an FRD-related
effect (see below).  However, in no case does the reference cable have
higher transmission than the best-transmission measurement for the
SparsePak fibers. Moreover, the transmission appears to be somewhat
lower (3\%) for the reference cable fed at f/13.5 instead of f/6.3.
We conclude that transmission variations between reference and
SparsePak fibers is not significant.

In summary, the SparsePak fibers are red-optimized and deliver total
throughput consistent with manufacturer's specifications. The total
throughput rises above 80\% redwards of 5000\AA, reaches 90\% redwards
of 6500\AA, and peaks near 92\% at 8000\AA.

\subsection{Focal Ratio Degradation}

While the total transmission of SparsePak fibers is high, also
relevant for spectrograph performance is the effective output focal
ratio of the fibers. A telescope delivers a converging (conical)
fiber-input beam, with constant surface-brightness cross-section and
square edges in the far-field for a point source. Telescope
obstructions (e.g., secondary and tertiary mirrors) make the beam
profile annular, still with constant surface-brightness within the
far-field annulus. The effect of fiber micro-fractures or micro-bends
(see e.g., Carrasco \& Parry, 1994) scatters or redirects the incident
light such that the output focal ratio is faster and the beam profile
softer (a beam cross-section no longer has constant surface brightness
and the edges are soft even in the far field). This is FRD.

Fibers, then, degrade the input beam by radial scrambling, and hence
increase entropy (they also provide complete azimuthal scrambling, but
this is unimportant here). The information lost can only be recovered
at additional cost (e.g., larger optics at the output end of the
fiber). One measure of this signal degradation is to measure the
output focal ratio containing some fixed {\it fraction} of the total
transmitted flux,, i.e. the encircled energy (EE). A perusal of the
literature (e.g. Barden, Ramsey \& Truax 1981) indicates the specific
choice of fiducial flux fraction is arbitrary. Carrasco \& Parry
(1994) prefer to parametrize fiber FRD by a more fundamental parameter
which characterizes their adopted micro-bending model. To the extent
that the model is correct, this has the strong advantage of being much
more general by enabling measurements of a given fiber to be used to
characterize the FRD performance of similar fibers of different
lengths. Here we measure the full input and output beam profile, from
which any index may be extracted.

In our test bench, we used f/4.5, f/6.3, and f/13.5 beams for the
intermediate focus which feeds the fibers in ``fiber'' mode.  These
focal ratios are the input beams produced, respectively by the
Hobby-Eberly Telescope (HET) spherical aberration corrector (which
feeds the Fiber Instrument Feed), the WIYN Nasmyth imaging port, and
the WIYN Modified Cassegrain port. We did not simulate, however, any
of the central obstructions in these systems.  Central obscurations
will steepen profiles of EE vs $f$-ratio for a pure imaging
system. For example, on WIYN the central obstruction is 17.1\%, which
is equivalent to f/15.3 relative to the f/6.3 beam at the Nasmyth
focus; no light is contained in the far-field within input cones
slower than f/15.3. The effects of FRD will be to scatter light into
this slow cone, as well as into a cone faster than f/6.3. However,
since we are interested primarily in the effect at small $f$-ratio,
the effects of the central obstruction will only be important if the
radial scrambling is gross. This is not the case.  Here we report the
results for the f/6.3 beam with the SparsePak and reference fibers.

Figure 6 shows that the SparsePak fibers have a wide range of output
beam profiles when all are fed with the same (f/6.3) input beam. As a
check on the quality of the imaging system, we also measured the beam
profile of the ``straight through'' system. We find the latter profile
is close to the ideal case of a constant surface-brightness (perfect)
beam, except near f/6.3, where there is a little droop indicating some
softness in our profile edges. However, the SparsePak fiber output
beams are so substantially aberrated in comparison with the ``straight
through'' beam, the imperfections in the optical system are
second-order effects. What is significant to note is that the
reference cable has an output beam profile very similar to the
``straight through'' system, i.e., the FRD in the reference cable is
very low. For example, the reference cable output beam contains over
90\% of its signal within f/6.3 (the input $f$-ratio), whereas the
mean SparsePak fiber contains only 67\% of its signal within this same
$f$-ratio.

\subsubsection{Wavelength Dependence}

We have checked that there is no {\it significant} wavelength
dependence to FRD. Figure 7 shows measurements of output EE at f/6.3
as a function of wavelength between 470-800 nm for four representative
SparsePak fibers. The variations between fibers at a given wavelength
is due to other effects, which we address in the following section.

There is some evidence for a {\it modest} FRD increase in the red for
the two fibers with the largest FRD, but no evidence for this effect
for the fiber with the smallest FRD. This is qualitatively consistent
with the micro-bend model adopted by Carrasco \& Parry (1994) which
predicts the broadening width of a collimated beam at large incidence
angles is proportional to $\lambda$, e.g., a factor of 1.7 between 470
nm and 800 nm. The effect should become larger when the overall
amplitude of the FRD increases.  A quantitative test of the wavelength
dependence of their model requires more precise measurements, and is
worthy of future pursuit: Carrasco \& Parry's (1994) direct
measurements of the broadening width was a factor of two short of the
model predictions. Other work by Schmoll, Roth \& Laux (2002) indicate
there is little wavelength-dependence to FRD, and cite other
theoretical work which predicts that there should be no wavelength
dependence. Clearly this issue is not resolved. For our purposes, FRD
wavelength-dependence, if it is real, amounts to less than or of order
a few \% variation in the indices we discuss next.

\subsection{Implications for future cable and spectrograph design}

We explored possible causes of the wide range in FRD for the SparsePak
fibers seen in Figures 6 and 7. The two likely causes, we believed,
would be slit position (due to the systematically changing radius of
curvature of the fibers along the slit), and array position, due,
possibly, to edge effects. Figures 8 and 9 show, respectively, the
$f$-ratio at fixed EE and relative EE at fixed $f$-ratio as a function of fiber
position along the slit.  These figures demonstrate, indeed, these two
fiber attributes explain essentially {\it all} of the profile variance
for the SparsePak fibers. 

The first-order effect is slit position: FRD is greatest for fibers
with the lowest number which are at the ``top'' of the slit where the
curvature in the foot is greatest. The radius of curvature of the
fibers goes from 82.6mm (for fiber \#1 at the top of the slit) to
178mm (for fiber \#82 at the bottom of the slit).  We conclude that
there would be substantial improvement (decrease!) in the FRD if the
fiber foot were straightened somewhat. The amount of straightening
needed is probably slight given the observed fact that the bottom
fibers (least bent) have FRD properties that nearly converge with the
Reference Fiber. Based on extrapolating the trend of EE50 with radius
of curvature for the SparsePak fibers to the reference cable, we would
recommend a minimum radius of curvature of 240mm for 500$\mu$m
fibers. It is not known if these FRD effects are present for the WIYN
cables; these have smaller fibers, which are more
flexible. Verification will await on-telescope measurements of these
thinner fibers.

It also appears that fibers within 1 fiber from the edge of the array
suffer a second-order increase in FRD relative to fibers at comparable
slit position but more centrally located within the fiber head.  As we
discuss in the Appendix, the process of releasing the head from the mold
undoubtedly induces stress on the edge fibers.  For this reason we
introduced a single layer of short, packing fibers around the entire
array. Clearly one layer was not enough. However, as seen in Figures
8 and 9, there is no evidence that fibers within 2-4 fibers depth
from the edge of the array behave any differently than more centrally
located fibers. We surmise, therefore, that just one more layer (2
layers total) of edge (buffer) fibers would have been sufficient to
have prevented this second-order enhancement of the FRD.  We note that
we have not proven the added stress originates from the mold-release
phase. The increased FRD could also stem from the asymmetric
distribution of pressure introduced from curing (contracting) epoxy
for edge fibers, or from the head-mount clamp. Whatever the cause, we
suspect that this is a generic result, independent of fiber diameter
for FFUs manufactured with a similar glue and press. To be safe, we
would recommend a minimum of 2 fiber layers of 500$\mu$m fibers, or
the corresponding number of fibers of different diameter to make at
least a 1.2mm buffer.

These results have ramifications for the throughput of the Bench
Spectrograph, and its improvement, as illustrated in Figures 9 and
10. Currently an f/6.3 beam is input to the fibers, and is then
matched to a spectrograph designed for a 152mm collimated beam.  The
initial design for the 4m Mayall telescope had 152mm and 203mm f/6.7
collimators (Barden \etal\ 1993a). The current Bench on WIYN has a 235mm
clear-aperture diameter parabolic collimator with focal length of 1021mm. This
collimator captures an f/4.2 for an on-axis fiber, but produces a
collimated beam substantially in excess of 152mm.  The effective f/6.7
beam is collimated into 152mm, but contains significantly less than the
full output of the fibers.
%
%
%

Adopting the mean curve for the laboratory FRD measurements for the
SparsePak cable, an f/6.7 beam contains only roughly 62\% of the fiber
output. The amount of light outside of the 152mm diameter beam in the
current system is in the range of 25\% to 55\%, and 38\% on
average. (Other curves for DensePak and Hydra cables, measured on the
telescope, indicate similar performance [P. Smith \& C. Conselice,
1998, private communication].)  If a faster collimator were used, the
fraction of enclosed light in a 152mm collimated beam would rise to
81, 88, and 95\% for 870mm, 790mm, 715mm focal lengths, respectively.
With no change in the camera, the resulting spectral resolution will
decrease with the increase in the magnification with the collimator
focal length. This penalty assumes that the slit is resolved. For
smaller fibers and set-ups with large anamorphic factors the
degradation in spectral resolution will be smaller.

Hence a trivial upgrade to the Bench Spectrograph, consisting of
inserting a faster, parabolic collimator, will improve the throughput
by 31, 42 and 53\%, with a loss in spectral resolution less than 18,
29, and 42\%, respectively for replacement collimator focal-length of
870mm, 790mm, 715mm. The decrease in spectral resolution will be less
severe for the smaller fiber diameters, since the current system is
under-sampled. Based on these results, we argue that a collimator with
focal length near 750mm would substantially improve the throughput of
the spectrograph for acceptable losses in spectral resolution. For
some applications, however, the loss in resolution will be
unacceptable. One of the better features of the Bench is its modular
and accessible design. This means that one could switch with relative
ease between one of several collimators.


\section{SUMMARY AND DISCUSSION}

In this paper we have presented the design, construction, and
laboratory measurements of the SparsePak FFU, a formatted fiber-optic
array designed to mate the Nasmyth imaging port to the Bench
Spectrograph on the WIYN 3.5-m telescope. A physical description of
the SparsePak array is found in \S 2 and 4, with a complete
description of the design and assembly contained in the Appendix. The
latter is included specifically to allow others to repeat this
relatively simple instrument-building process.

The primary scientific motivation for this array is to measure the
kinematics of stars and gas in nearby galaxy disks. Examples of
science capabilities are found in Paper II, and in several recent
studies using SparsePak to measure stellar velocity dispersions to
estimate dynamical mass and the radial dependence of the mass-to-light
ratio in spiral disks radius (Bershady \etal\ 2002, Verheijen \etal\
2003, 2004), the Tully-Fisher relation in barred spirals (Courteau \etal\
2003), and rotation curves and ionized gas kinematics in
low-surface-brightness galaxies (Swaters \etal\ 2003).

Here we have focused on the confluence of science objectives with
technical, observational, and physical constraints that shaped the
specific SparsePak design. The salient design drivers are the
balancing of trade-offs in spatial and spectral coverage and
resolution, plus an optimization for superior sky-subtraction. For
this latter purpose we have developed an analytic expression for the
optimum number of sky fibers in the background- and detector-limited
regimes. Placement of sky and source fibers along the spectrograph
slit are also important considerations in our design -- both for sky
subtraction and to offset the effects of vignetting within the
spectrograph. In Paper II we explore the sky-subtraction performance
of our design. The results presented here and in Paper II are
applicable to other fiber-fed spectrographs.

The SparsePak performance can be summarized in terms of its
instrumental grasp of 137 m$^2$ arcsec$^2$ per fiber, and 11,200 m$^2$
arcsec$^2$ overall. In more details, the cable has a throughput of
89-92\% redwards of 500 nm (92\% is the expected best-case for two
fused-silica--air interfaces). The cable throughput drops rapidly
below 500 nm. The overall system throughput (telescope plus cable plus
spectrograph) are measured and described in Paper II. The fibers
produce FRD which takes an input f/6.3 beam and degrades to f/5.7 to
f/4.1 at EE95. There is no wavelength dependence to the observed FRD
between 500 and 800 nm. The non-telecentricity error budget is below
0.1 deg (differential) at the telescope focal-plane, and does not
contribute significantly to the observed FRD. The range of FRD
depends, to first order, on a fiber's slit position (due to mechanical
curvature of spectrograph feed) and, to second order, on its position
in the head (due to edge stresses).  The implications are that future
fiber-arrays manufactured with the process described here should have
thicker buffers of inactive fibers (at least 2 fibers thick) on the
outside walls of the array, and straighter feeds.

There are several conclusion which may be drawn from our study. First,
we have shown that economical, high-performance optical cables are
possible to build which are relatively long in length.  The losses
within the fibers themselves are relatively small; in a fiber cable of
25m length the end-losses dominate. This means that significantly
longer cables can be constructed with high-throughput performance in
the red. This may be useful for connecting spectrographs to multiple
telescopes at a given site.

Second, our laboratory measurements indicate the promise of
significantly lower levels of FRD with modest modifications to our
fiber-head design construction and closer attention to the bending
geometry near the spectrograph termination -- even for very large (500
$\mu$m) fibers. This optimism should be tempered by the fact that all
other existing cables on the Bench Spectrograph exhibit comparable FRD
to SparsePak. However, it is unclear whether the other Bench fiber
feeds have been optimized to minimize fiber stress.


Finally, within the context of the current fiber cables feeding the
WIYN Bench Spectrograph, it is clear that much of the light is being
lost due to the injection of a beam faster than the spectrograph was
designed to handle. In Paper II we document how much of the light is
lost from the SparsePak cable due to geometric vignetting factors
within the Bench Spectrograph. Roughly speaking, a modest decrease in
the collimator focal-length would gain back factors of two in
throughput while at worst decreasing the spectral resolution by only
30\%. For many programs this trade of throughput for resolution is a
winning proposition. Moreover, the resolution losses are not even this
severe since the current system is under-sampled; higher resolution
can be regained for stellar surveys by using smaller fibers with
little light-loss in the telescope focal plane -- thanks to WIYN's
excellent image quality. Given the cost of running the WIYN telescope
and the number of nights the Bench Spectrograph is in use, it would
seem unimaginable were such an upgrade not implemented in short order.

\acknowledgements

We wish to thank S. Barden, L. Engel, and D. Sawyer for consultation
on IFU and cable design; C. Corson, D. Harmer, and G. Jacoby for
making this a reality at WIYN; D. Bucholtz, without whom the many
critical details of SparsePak would have languished; S. Buckley, and
D. Hoffman for their excellence in instrument making; and many
helping-hands at UW to pull the cable out of the stairwell (several
times). We also thank R. Swaters for uncovering an error in our
assumed plate scale, adopted from the DensePak web-page prior to
2001. Support for this project is from NSF
AST/ATI-9618849. M.A.B. also acknowledges support from NSF grant
AST-9970780 and the UW Grad School.

\begin{appendix}

\section{TECHNICAL DESCRIPTION AND FABRICATION}

\subsection{Cabling: Design}

Each of the 82 science fibers, 25.37m in length, is housed in a 24.5m
cable, terminated at the telescope end by a ``head'', and at the
spectrograph end by a curved ``foot.''  The head and foot,
respectively, contain 0.32m and 0.55m of fiber.  The lay-out of the
fibers in the head is illustrated in Figure 2, described in \S 2, and
motivated in \S 3. The optical fiber is a single draw of
Polymicro\footnote{Polymicro Technologies Incorporated, 18019 North
25th St, Phoenix, AZ 85023-1200, (602) 375-4100.} ultra-low OH$^-$,
multi-mode, step-index fiber with pure fused-silica core, doped silica
clad, and polyimide buffer with core:clad:buffer diameters of
500:550:590 $\mu$m, respectively. Our order was for 2.3km of fiber, at
a cost of $\sim$\$35,000 in 1997.

The cable consists of a ``core'' of 82 opaque-black 18-gauge Teflon
tubes\footnote{Zeus, PO Box 2167, 620 Magnolia Street, Orangeburg, SC
2911-2167, (803) 533-5694.} (one per fiber), surrounded by standard
PVC tubing, and finally covered with stainless-steel, interlocked,
flexible conduit\footnote{McMaster-Carr Supply Co., PO Box 94930,
Cleveland, OH 44101-4930, (330) 995-5500.} with a 5.72cm inner
diameter (ID), 6.35cm outer-diameter (OD), and a minimum bend radius
larger than 19cm. This multi-layered cable design provides a
well-supported, smooth conduit and protects the stiff, 500 $\mu$m
fibers from breakage and over-bending.


Because the bundled Teflon tubes have a much smaller diameter (roughly
1.90cm) than the exterior conduit, we fed the Teflon into PVC tubing
with 2.54cm ID, and jointed the conduit every 1.8m with connectors
just under the exterior (stainless steel) conduit ID. This
minimizes sagging, and hence differential path lengths between the
exterior cable, the Teflon, and ultimately the fibers. The aim
is to minimize a potential source of fiber stress.

The PVC is attached rigidly to the exterior stainless conduit at both
ends via set-screws pressing on metal connectors attached to the
PVC. The set-screws are threaded through aluminum collars which also
serve to terminate the stainless conduit and provide a mounting
surface for the fiber head and foot. Two other collars join together
three $\sim$7.6m lengths of stainless conduit, and allow cable access. Due
to the natural curvature in the PVC, the segmented PVC cable is quite
elastic and conveniently acts to pull the cable collars tight against
the end of the conduit even in the absence of rigid attachment.

The Teflon is terminated slightly above and below the conduit within
interface modules joining the cable to the head and foot assemblies.
At the head-end of the cable, Teflon extends up an additional 10cm
and terminates in an aluminum anti-rotation collar which is rigidly
attached to the cable termination collar via a 5.08cm OD cylindrical
aluminum flange. This collar consists of a circular array of 82 holes
through which the Teflon is pulled with little clearance. The Teflon
ends are flared with a soldering iron, and a short (1cm) piece of
shrink-wrap is placed just behind the flare to form a collar. This
prevents the Teflon from sliding back through the holes in the
anti-rotation collar, but it does not prevent the Teflon from pushing
up.

The foot-end of the cable connects to a replication of a standard
NOAO/WIYN Bench Spectrograph ``cable interface.''  Teflon terminates
0.19m below the cable at the end of the interface in a 3-element shear
clamp with holes arranged in a staggered, rectangular pattern (35
holes in length and 2 or 3 holes in width). This serves to translate
the round fiber bundle into a linear slit and anchor the Teflon
against movement in either direction.

\subsection{Cabling: Assembly}

Assembly took place in public space within the Astronomy Department at
the University of Wisconsin. (For obvious reasons, this effort would
have benefited substantially from a dedicated space.) Pre-cut Teflon
tubes were unrolled horizontally in a clean hallway on a packing paper
bed (to minimize dust), grouped in numbers of 6 or 7 with shrink-wrap,
and labeled.  At the telescope (``top'') end these groups were bundled
into a single, flared unit via a larger piece of shrink-wrap. The
flare and grab of the multiple layers of shrink-wrap was sufficient to
suspend the entire Teflon core vertically under its own weight in an
8-story stair-well. The PVC and stainless conduit were pulled up and
over the Teflon core, until the cable was hanging by the stainless
conduit with the flared Teflon bundle resting on the neck of the
terminal PVC connector.  The anti-rotation collar and flange (top-end)
and cable interface and sheer clamp (bottom-end) were then installed,
and the Teflon permanently locked into place. At this stage the
mapping between the fibers in the FFU head and slit were effectively
set (see \S 3.2.6 and 3.3.2).

Fibers were individually measured via a calibrated transfer spool, cut
using a ruby cleaver, and fed from the top.  Because the 500 $\mu$m
fibers are very stiff and the Teflon was hanging vertically and
relatively unentwined, it was easy to install the fiber.  We required
a J-bend at the bottom end of the cable to keep some fibers from
slipping all the way through. The feeding process took 15 minutes per
fiber for preparation (length measurement and cleaving), and the same
for installation. The two tasks were done in parallel. Once installed,
we identified and labeled the fiber at both ends of the cable,
checking at the same time that there were no breakages. The cable was
then hauled out of the stairwell and placed into its traveling box for
movement to the polishing and optics lab.

\subsection{Fiber Head: Assembly}

The fibers in the fully assembled cable were glued into an
``integral'' head using a pressing jig, or mold. The jig design
(Barden, private communication) consists of a precision-cut, U-shaped
channel, 76mm in length, made from three walls, one of which is
precision-actuated on an under-cut block with a micrometer for
repeatable width adjustment (see Figure 11). Before gluing, the jig
channel is thoroughly sprayed with dry lubricant mold-release
agent\footnote{Miller-Stephenson Chemical Co., Inc., 55 Backus Ave.,
Danbury, CT 06810 USA, (203) 743-4447.}. The gluing process begins
with pre-setting the channel width and machining a ``tamping'' tool to
precisely fit this width. Short, packing fibers, 50.8mm in length were
precut and laid down in rows, interspersed with the appropriate long
fibers. Each row is wicked with glue as it is placed into the jig.  As
the entire array is assembled, the tamping tool is used to press the
array into its tightly packed form, and to squeeze out excess
glue. There are a total of 367 packing fibers in the head. The face of
the array, which at this stage consists of ruby-cut fiber edges is
monitored via a mirror and microscope to ensure correct positioning of
the long, science fibers. Since our emphasis when gluing the head was
to make sure fibers were well-seated with minimal flaring, we did not
force fiber-ends to be even, and hence the exact termination length
differs by $\pm$1-2mm from fiber to fiber. Given the limited
depth-of-field of the microscope the assessment of fiber position was
aided by back-lighting science fibers.

The entire process of laying down precision-placed rows such that the
bundle has little-to-no flare in the out-of-channel dimension requires
skill and finesse which can only be accomplished through practice.
Very small angular deviations from telecentricity can lead to
substantial effective FRD, as illustrated in Figure 12 (see also Wynne
1993). Hence, assuring that there is no flaring of the bundle is of
utmost importance.

We use EPO-TEK 354\footnote{Epoxy Technology Inc., 14 Fortune Drive,
Billerica, MA 01821, (508) 667-3805.}, a temperature-curing epoxy. The
heads were cured in-situ within the pressing jig with a heat lamp, and
then released. In testing we found that despite liberal
application of mold-release, occasionally edge fibers were damaged or
fractured. Consequently we added a minimum of 1 row or column of
short, packing fibers to surround, or buffer, all elements of the
science fiber arrays. As we show in \S 6, we have found edge-effects
in the SparsePak array. On this basis we would now advocate a minimum
of 2, and ideally at least 3 rows of packing fibers to act as a
buffer.

\subsection{Fiber Head: Polishing}

Fiber head polishing was done using an Ultrapol 1200 polishing/lapping
machine\footnote{Ultra Tech, 1025 E. Chestnut Ave., Santa Anna, CA
92701-6491, (714) 542-0608.} with 15.25cm circular lapping
disks.\footnote{Moyco Industries, Inc., Corporate Offices \&
Ultralap/Abrasives Division, 200 Commerce Dr., Montgomeryville, PA
18936, (800) 331-8837.} The polisher consists of a horizontal,
rotating, aluminum platen, lubricant tub, and polishing head with
precision 3-axis positioner and oscillator.  The lubricant of choice
is distilled water. Many of the features on the costly positioner were
unused, and in hind-sight a custom made positioner would have been
optimal for application with the heavier fiber arrays.  In all cases
fibers were attached via custom-made, aluminum adapters that attached
directly to the polishing head. The fiber head adapter consisted of
two 6.35mm thick L-shaped brackets which screwed together diagonally
to place pressure on all four walls. To ensure the polished surface
was orthogonal to the fiber length, support walls were connected to
the L-brackets (see Figure 13).

Initial material was removed with course grit before we began true
polishing. The polishing process consists of descending in grit size
from 60, 40, 30, 15, 10, 5, 1 (silicon carbide) and finally 0.5 $\mu$m
(aluminum oxide). For single fibers or small FFU heads made from
thinner fibers, the oscillator could be used, greatly facilitating
high-grade polishing. A single fiber can take as little as an hour to
polish, while a small FFU polishes in roughly 1-2 days. SparsePak head
polishing took several weeks due to the large area and quantity of
material removed and the mechanical difficulty of manipulating the
cable.

For example, because the cable was stiff and heavy, it had to be
specially suspended and supported during polishing by a combination of
horse and camera tripod. A special brace also was made to connect
directly the polishing head adapter to the cable termination, to
ensure fibers were not crushed (Figure 13). Given these modifications,
it was time consuming to remove and inspect the SparsePak head in
mid-polish, and nearly impossible to replace the head at exactly the
same angle. Hence small facets were introduced, and a ``final''
polish, while yielding excellent luster, still left several, discrete
scratches on the head surface visible at back-lit, oblique angles at
10$\times$ magnification or higher. Horizontal polishing with the
existing mount hardware was unsatisfactory for SparsePak.

We therefore developed a vertical hand-polishing tool using two sets
of precision ball slides\footnote{\#E-4, Del Tron Precision, Inc., 5
Trowbridge Dr., PO Box 505, Bethel, CT 06801, (800) 245-5013.} to
provide an x-y stage with 110mm of diagonal travel (see the SparsePak
web-site). A hand-positioned, removable plunger with a 70mm circular
polishing surface allowed us to polish, remove, and inspect the
SparsePak fiber-head at regular and frequent intervals.  (Lubricant is
dropped vertically through the system while polishing.)  With this
polisher we were able to ensure perpendicularity of the polishing
surface, remove remaining facets, and achieve a superior polish on the
SparsePak surface. Despite these improvements, the final Sparsepak
head face shows small micro-scratches on fibers 4, 16, 17, 24, 36, 40,
72, even though the over-all luster is high. Based on our lab
measurements described in \S 6, there is no evidence that these
features diminish the throughput or degrade the output focal ratio by
a measurable amount.

Our experience with EPO-TEK 354 is that while its wicking
properties are good, it does not polish as well as, e.g., Norland 68
UV curing epoxy\footnote{Norland Products Inc., PO Box 637-T,
Cranbury, NJ, 08512, (609) 395-1966.}. However, given the thickness of
the SparsePak fiber head, a heat-curing epoxy was essential.

\subsection{Fiber Head: Mount}

SparsePak is designed to be swapped into and out of the WIYN Nasmyth
imaging port (IAS) on a regular basis. The terminal mechanical element
in the SparsePak cable, the head mount, serves to grip the fiber head
and provide a mounting surface. Reported here are salient details
required to provide a rigid and robust mounting mechanism compatible
with existing interface hardware on the telescope.

A mounting box (WIFOE, named after the WIYN Fiber-Optic Echelle),
developed by K. Honneycutt and collaborators for a single fiber-optic
feed, is the mechanical assembly to which the SparsePak and DensePak
attach during operation. WIFOE contains optics for feeding line-lamps
to the focal plane, and for simultaneously viewing the back-lit fiber
face and focal-plane image from the telescope.  The WIFOE port
requires a 25.4mm OD tube of minimum length 44.4mm, permitting a 17mm
maximum diameter for the fiber array in order to maintain mechanical
rigidity in the surrounding head mount. The SparePak fiber head
diagonal dimension is at this limit.

The SparsePak fiber head mount (Figure 14) was designed to ensure that
the array was held at the proper telecentric angle if held rigidly by
the WIFOE port. To do so we cut a 50.8mm-long channel in rectangular
aluminum stock. The channel width matches the SparsePak fiber-head
width of 11.7mm, with little ($\sim 25 \mu$m) clearance. The stock was
then turned down to a 76.2mm tube with a 25.4mm diameter, and a 6.35mm
thick flange with a 50.8mm OD for mounting to the rest of the
cable. Three half clam-shell clamps cover the channel. These units
serve to give the tube a nearly complete circular cross-section, while
at the same time providing the clamp mechanism for the array. The
array was held with a uniform-pressure rectangular clamp seated below
the clam-shells, actuated by three set-screws in the clam-shells. The
fiber head is padded on both top and bottom by a thin rubber gasket
38.1mm in length. In this way, substantial force is exerted on grabbing
the fiber without substantially stressing the glass.

To protect the fiber-head face during installation and removal, the
fiber head terminates roughly 7.6mm back from the end of the mounting
tube.  This last portion of the mounting tube was beveled, ringed,
sand-blasted, and carefully flocked to avoid vignetting of the input
beam on the edge fibers, while minimizing glints and scattered
light. Figure 15 is an image of the front face of the final, assembled
fiber head and mount.

The SparsePak fiber head-mount is fastened in the WIFOE
entrance-collar with a thumb-screw that sets into a detent. For a
light-weight cable, such as that containing a single fiber, this
fastener is adequate both to hold and retain the cable robustly at the
correct telecentric angle as the IAS rotates during observations.
Because of SparsePak's considerable weight and rigidity, it was
necessary to create a removable support flange. The flange attaches to
the cable $\sim$33cm back from the port collar, and consists of a
rigid, closed brace with a three-leg attachment to the WIFOE and
IAS. This modification eliminated remaining flexure to levels below
detection, and can also be used with DensePak.

The overall accuracy of the telecentric positioning of the fiber head
is estimated to be better than $0.2^\circ$, based on the quadrature
sum of our estimates for (a) the accuracy of the positioning of the
fiber head within the head mount ($<0.15^\circ$); and (b) the accuracy
of the positioning of the head-mount within the WIFOE collar
($<0.15^\circ$). Both of these estimates are based on the known
mechanical tolerances of the mounting hardware. No accommodation in
this estimate is made for any non-uniformities in the Nasmyth port
mounts and WIFOE. The resulting increase in FRD, as illustrated in
Figure 12, is under 3\%.

\subsection{Fiber Slit: Mount, Assembly and Polishing}

The slit assembly, containing all elements after the shear-clamp,
consists of a 90$^\circ$ curved foot (Figure 16), where bare fibers
make bends with radii between 82.6mm and 178mm, a slit block, and
``toes'' for filters and a slit-narrowing mask. These are standard
components required to integrate with the existing spectrograph mount.
As shown in \S 6, the curvature in this foot is problematic, re-design
of which, while beyond the scope and budget of this project, presents
a clear upgrade path. However, we did make cost- and
performance-effective modifications to the toe and slit-block design.

We determined the standard toe design produced substantial vignetting
in the dimension transverse to the slit for on-axis beams faster than
f/5.7, and for beams faster f/7.1 emanating from the edge of a
500$\mu$m fiber.  There is also vignetting in the dimension parallel
to the slit for fibers near the ends of the slit; a slit-edge fiber in
SparsePak is vignetted for rays exiting faster than f/5.6. (For other
feeds the vignetting for edge fibers is more severe because of larger
slit-lengths -- 76.4mm instead of 73.6mm. This adds appreciably to
their over-all ``slit-function.'') Given the f/6.3 input beam from the
telescope, the output $f$-ratio of the fibers is expected to be
substantially faster than f/6 due to FRD. Hence the SparsePak toes'
chamber-baffles were enlarged and the last (4th) chamber removed. This
enables an output beam at f/4 at the end of the slit for a 500$\mu$m
fiber to exit unvignetted from the feed; provides room for
simultaneous use of one interference and one glass filter; and is
compatible with the existing mechanized filter-insertion mechanism on
the Bench.

The standard slit-block consists of two clamp plates, one of which
contains precision-machined v-grooves to locate micro-tubes, each
containing a fiber. Such machining is difficult and expensive.  We
found that for 600$\mu$m OD fibers, 20-gauge, thin-wall
stainless-steel micro-tubes (31.8mm in length, 647 $\mu$m ID and 902
$\mu$m OD)\footnote{Connecticut Hypodermics, 519 Main Street,
Yalesville, CT 06492, (203) 284-1520.}  provided an outstanding way to
hold and separate the fibers if the tubes were tightly packed like the
fibers in a row if the FFU head.  A packed row of 82 tubes yields a
fiber-to-fiber edge separation of over 400 $\mu$m, and a total slit
length of 73.6mm from active fiber edge to edge. The glue thickness is
inconsequential since the micro-tube were bonded in our pressing jig;
the glue bonded in the interstitial regions. Our adopted slit-block
consisted of three rows of micro-tubes glued together for mechanical
strength, pressed into a stainless steel holder, and pinned for
precision alignment to the end of the fiber foot (see Figure 16).  We
estimate that the overall alignment error of the fibers with respect
to the optical mounting axis of the fiber foot is under 0.2$^\circ$.

The fibers themselves were bent, cut to final length, and then glued
into the micro-tubes with only a bead of Norland 68 UV curing epoxy at
the end of the tubes, and cured using a commercial black-light.  The
entire slit-block, attached to the fiber foot, was mounted on our
Ultrapol lapping machine with a special adapter. It polished well, and
did not require additional hand-polishing. This concluded the
SparsePak cable manufacture.

\subsection{Total Effort}

For reference for future efforts: The cable construction and fiber
installation took $\sim$320 person-hours (2 people at any given,
except for hauling the cable -- a half-hour process requiring 4
people), and one month of shop-time for fittings and termination
hardware. Polishing took an additional person-month plus one week of
shop time. These totals do not include the more substantial effort to
develop the assembly, gluing and polishing techniques, nor does it
include time for laboratory calibration measurements described in \S
6.

\end{appendix}

\clearpage

\begin{figure}
\figurenum{1}
\plotfiddle{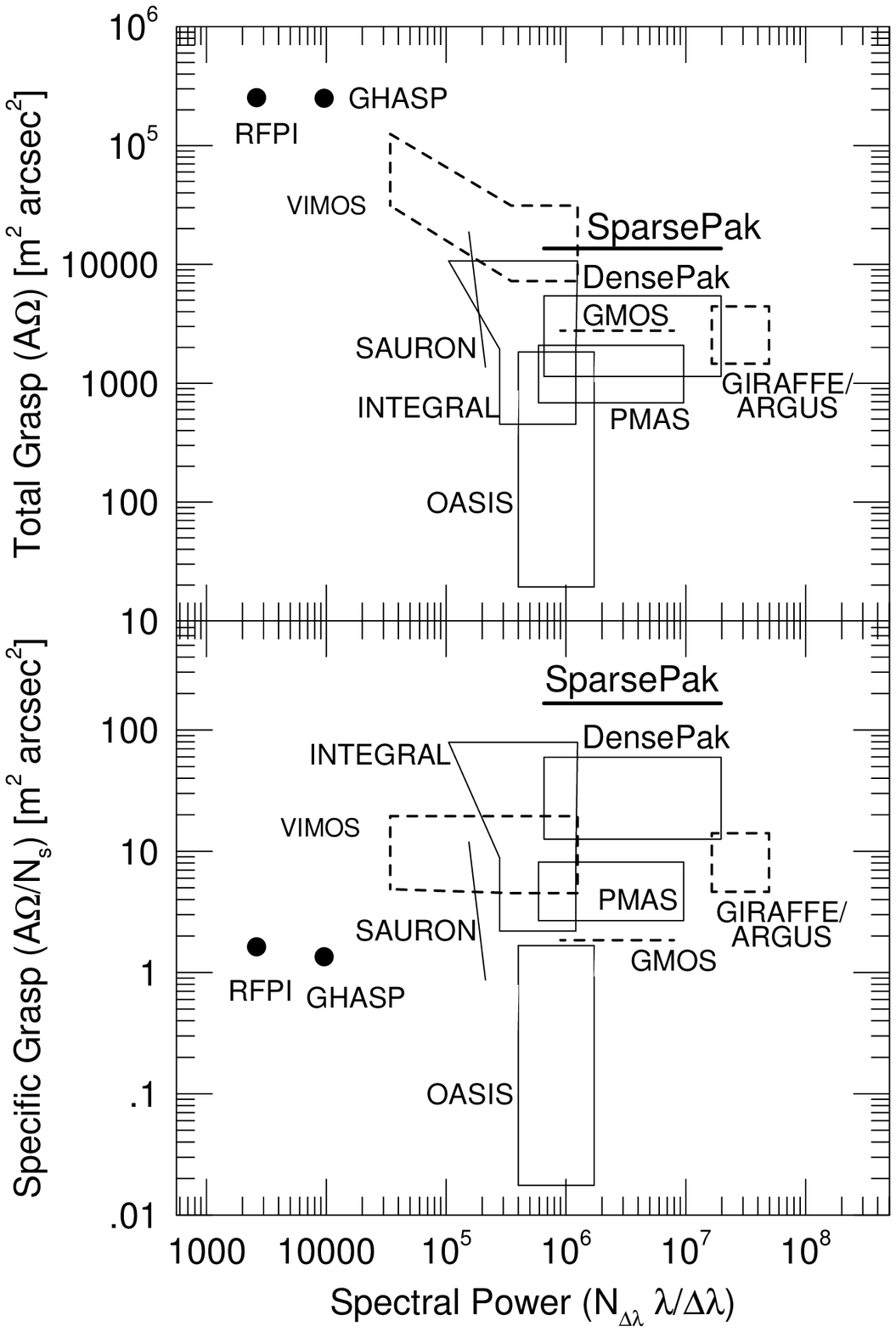}{5.5in}{0}{80}{80}{-230}{-75}
\caption{ Grasp versus spectral power for a
current suite of two-dimensional spectroscopic systems including
SparsePak (see text). The total grasp is defined as the product of
area $\times$ solid-angle (A$\Omega$). The specific grasp is the grasp
per spatial resolution element (in the case of SparsePak, this is per
fiber); N$_s$ is the number of spatial resolution elements.  The
spectral power is defined as the product of the spectral resolution,
$R=\lambda/\Delta\lambda$, times the number of spectral resolution
elements, N$_{\Delta\lambda}$. Spectographic instruments on 8m-class
telescopes are shown as dashed lines (Gemini/GMOS, VLT/VIMOS, and
VLT/ARGUS); spectographic instruments on 4m-class instruments are
shown as solid lines (WHT/SAURON and INTEGRAL, CFHT/OASIS, Calar
Alto/PMAS, and WIYN/DensePak and SparsePak); Fabry-Perot instruments
(GHASP and RFPI) are shown as filled circles. The variations in the
shapes of covered parameter space depends on how a given instrument
achieves a range of spectral resolution and spatial sampling, i.e.,
through changes in gratings, slit-widths, or both.  Note the unique
location of SparsPak in these diagrams.}
\end{figure}

\clearpage

\begin{figure}
\figurenum{2}
\plotfiddle{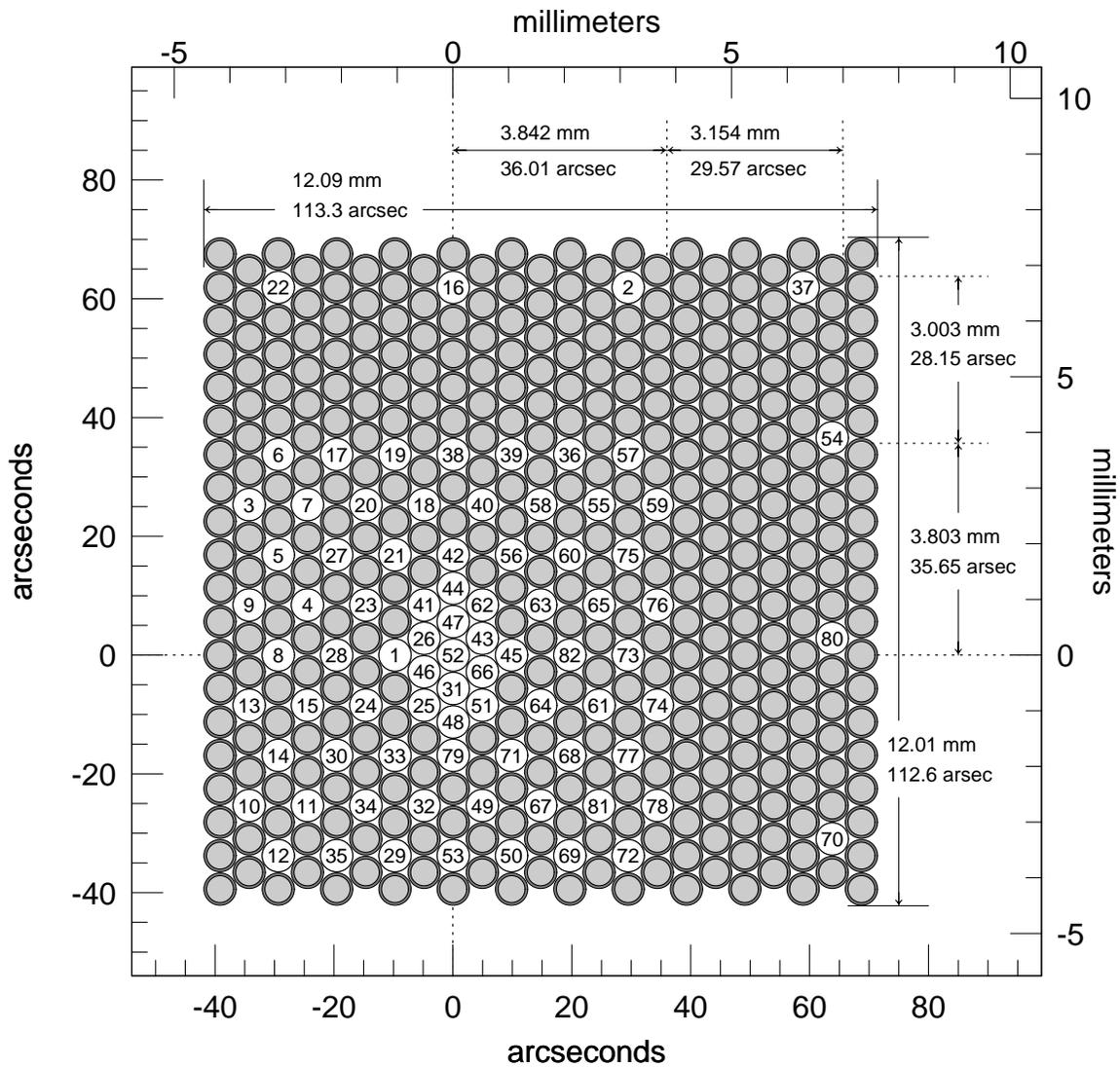}{5.5in}{0}{80}{80}{-240}{-75}
\caption{Astrometric diagram of the SparsePak
array head at the WIYN Nasmyth imaging port (bare RC focus). Active
(science) fibers are numbered according to their position in the slit.
The relative size of the fiber core and the core plus buffer is shown
to scale for the un-numbered, inactive, or buffer fibers. The active
fibers have the same geometry. On-sky orientation at the WIYN IAS port
with zero rotator offset places N upward and E to the left (also as
viewed in the WIFOE slit-viewing camera). Physical measurements were
made in our lab (as described in the text); angular dimensions are
based on these measurements using the nominal WIYN bare-RC
imaging-port plate scale of 9.374 arcsec/mm.}
\end{figure}

\clearpage

\begin{figure}
\figurenum{3}
\plotfiddle{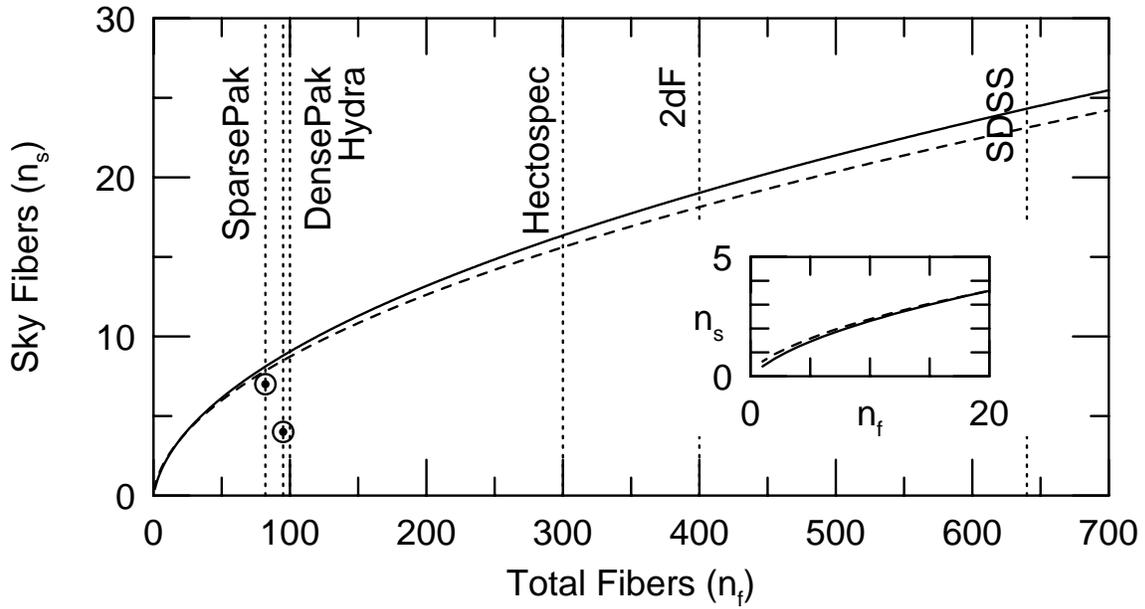}{4.5in}{0}{90}{90}{-280}{-150}
\caption{ Optimum number of fibers dedicated to
sky (n$_s$) is plotted as a function of the total number of
spectrograph fibers (n$_f$). This assumes that sky backgrounds and
their subtraction contribute only to random (shot) noise,
i.e. systematic errors are not considered in this model.
Background-limited measurements are indicated by the solid line;
detector-limited measurements by the dashed line. An insert
illustrates the behavior at small n$_f$. The two points represent the
number of sky fibers allocated for SparsePak and DensePak. For
SparsePak, the optimum number of sky fibers is $\sim$8, while for
DensePak the optimum number is number; the allocated numbers are 7 and
4, respectively. However, the survey merit function defined in the
text is a weak function of n$_s$. The relevant number of sky fibers
for several other survey spectrographs are indicated.}
\end{figure}

\clearpage


\begin{figure}
\figurenum{4}
\epsscale{0.9}
\plotone{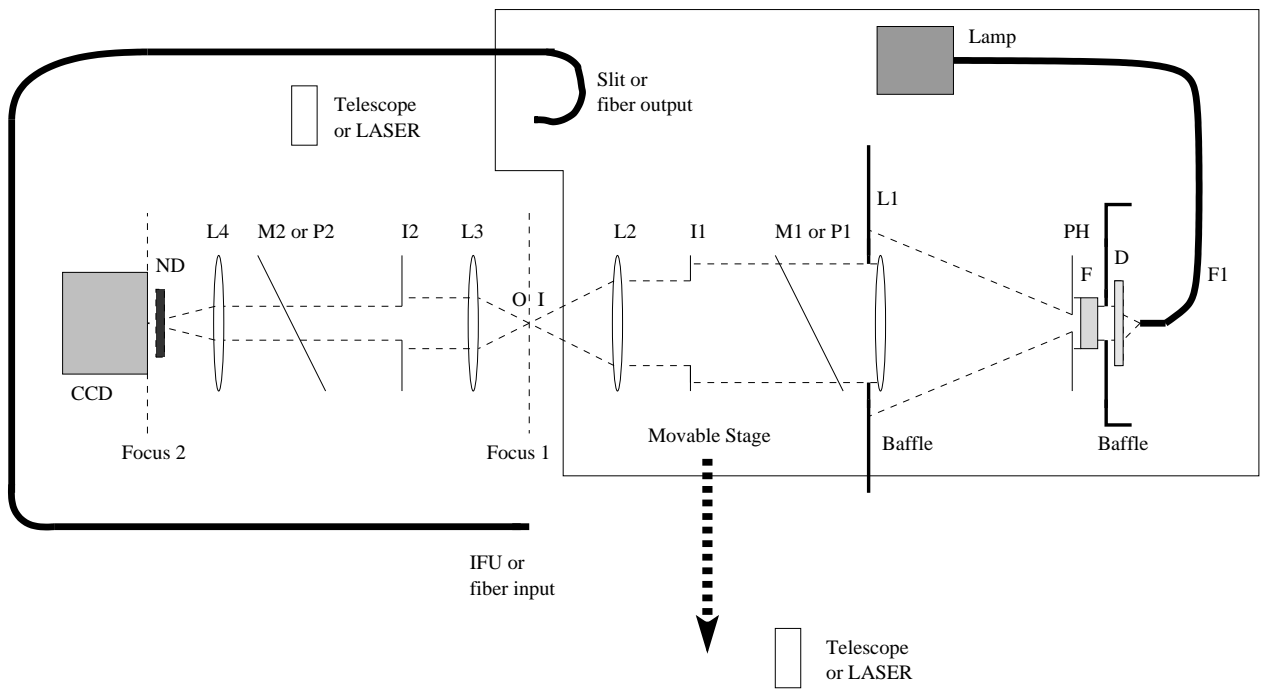}
\caption{ Schematic diagram of our test-bench for characterizing the
throughput and FRD of the optical fibers. The setup consists of a
double re-imaging system. The first system (right side, moving left
from lamp through camera optic L2), is on a translation stage such
that it can either feed the second system directly or via a fiber
feed. The setup is illustrated for the direct feed, or ``beam mode.''
The ``fiber mode'' is accomplished by moving the translation stage
such that the input (I) of Focus 1 from the first system is placed on
the fiber input or IFU (refer to labeled points). This translation
also moves the fiber output or Slit into the former position of Focus
1 to create the output (O) which feeds the second system (refer to labeled
points). The second system (left side, moving left from collimator
optical L3 to the CCD detector) is fixed.  Mirrors and pellicles
(P1/M1 and P2/M2) are placed into the collimated beams for alignment
purposes. Irises\ (I1 and I2) are for controlling the input and output
$f$-ratio, respectively.  Details of the light source, diffuser (D),
filter (F), and pin-hole (PH) setup, as well as the CCD and
neutral-density filters (ND) are given in the text.}
\end{figure}

\clearpage


\begin{figure}
\figurenum{5}
\plotfiddle{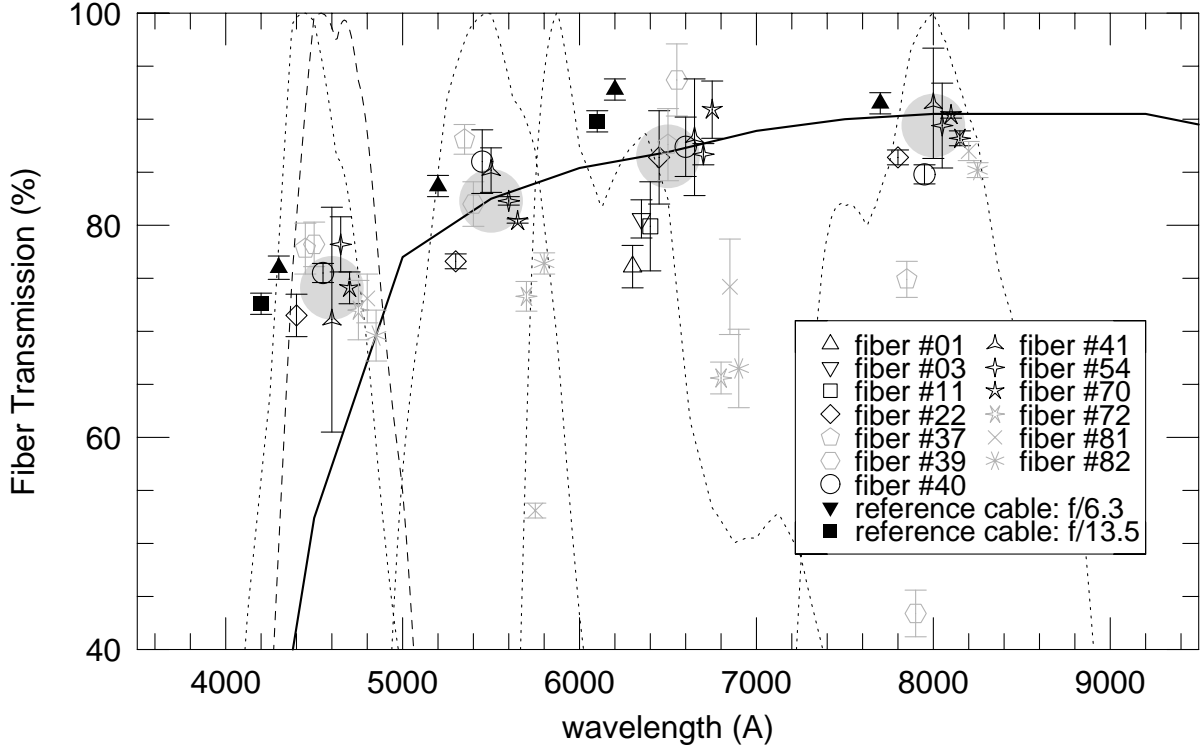}{5.5in}{0}{80}{80}{-240}{-75}
\caption{ Fiber transmission (throughput) of 13 of the 82 SparsePak
fibers and the SparsePak Reference Cable, measured using the
test-bench in our lab, as described in the text and illustrated in
Figure 13. Symbols denote different fibers, as identified in the key,
and are offset in wavelength within groups for presentation purposes.
Light symbols (gray instead of black) are fibers for which the
throughput variance across bands was larger than 15\%. These fibers
either had measurement-log notes indicating unsatisfactory setup of
the pin-hole image on the fiber face, or beam-mode measurement
variances indicating the lamp stability was poor. As such, these
measurements should be viewed as suspect, consistent with the
unusually high or low measured transmittance. The large, grey circles
are the median of the 8 good fiber measurements.  The solid curve is
the expected transmittance of 25.4m of Polymicro ultra-low OH-
fused-silica fiber, based on Polymicro's figures, combined with two
silica-air interfaces (3.43\% per interface). The dotted curves
represent the normalized broad-band filter transmission, convolved
with the CCD response function. From left to right: $B, V, R, I$.  For
$V,R,I$, note the excellent agreement between the measured and
expected values, the measurable difference between different
fibers. The $B$-band measurements appear high relative to
expectations. The dashed curve curve takes into account the effective
band-pass given the expected fiber transmittance. Not taken into
account is the spectrum of the light source. The likely effective
wavelength of the $B$-band measurements is 4900\AA.}
\end{figure}

\clearpage


\begin{figure}
\figurenum{6}
\plotfiddle{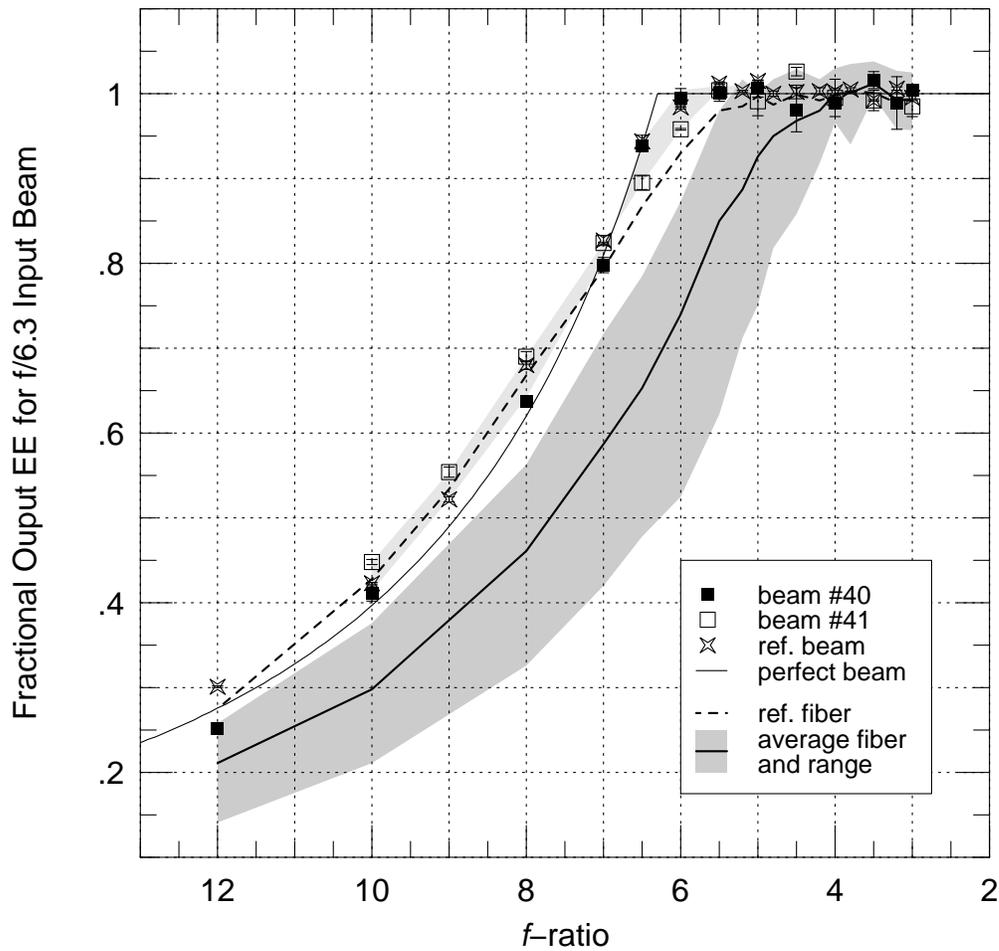}{5.5in}{0}{80}{80}{-240}{-75}
\caption{ Relative encircled energy (EE) as a function of output
$f$-ratio for an f/6.3 input-beam into SparsePak and Reference Cable
fibers. A perfect beam (with constant cross-sectional
surface-brightness) is represented by the thin, solid
line. Straight-through beam profiles (no fiber) measured when testing
three of the fibers (\#40, 41, and the reference cable) are shown to
illustrate the quality of the optical test-bench setup. The average
output beam profiles of the fibers is represented by the thick, solid
line; the grey shaded area represents the range for the 13 SparsePak
fibers measured.  {\it The range of FRD is real, and represents
systematic fiber-to-fiber differences due to variations in physical
conditions of the fibers} (see text). The reference cable fiber output
profile is the dashed curve. Reducing the curvature of the SparsePak
spectrograph-feed would dramatically improve the SparsePak fiber
profiles.}
\end{figure}

\clearpage


\begin{figure}
\figurenum{7}
\plotfiddle{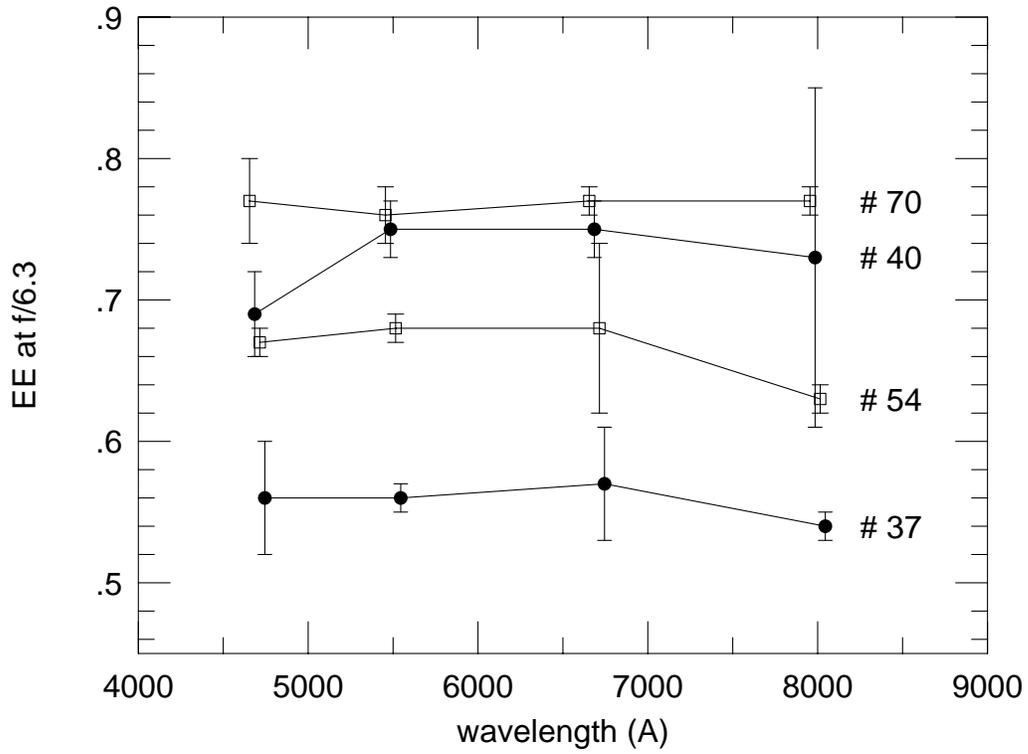}{5.5in}{0}{80}{80}{-240}{-75}
\caption{ Relative encircled energy (EE) at an output $f$-ratio of
f/6.3 as a function of wavelength for 4 of the SparsePak fibers
(labeled) fed with a f/6.3 input beam.  These represent four different
measures of FRD wavelength dependence for the 500 $\mu$m core
Polymicro ultra-low OH$^-$ fibers. Error-bars represent estimated
observational uncertainties. Wavelength offsets of points are for
presentation purposes.}
\end{figure}

\clearpage


\begin{figure}
\figurenum{8}
\plotfiddle{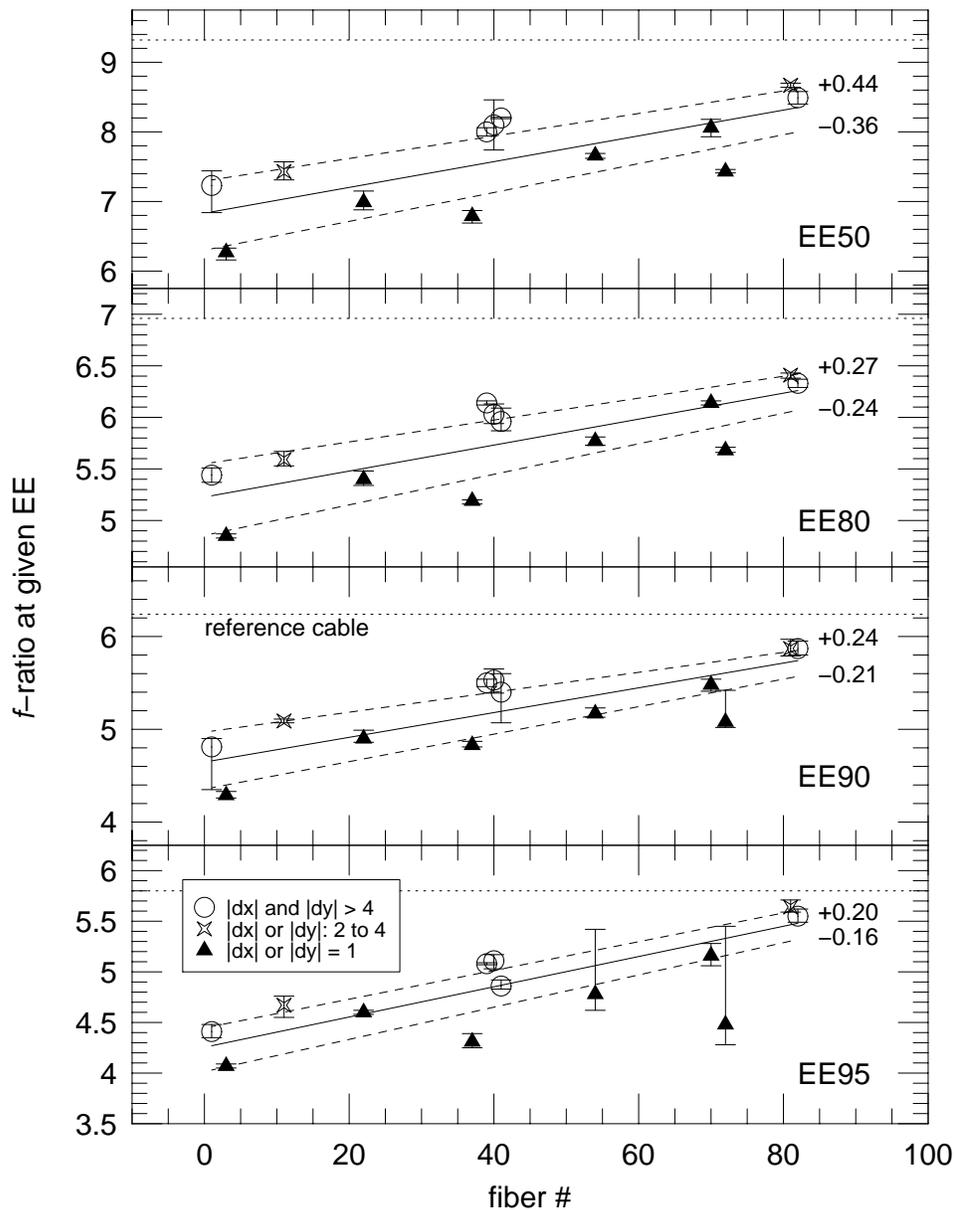}{6.0in}{0}{75}{75}{-240}{-50}
\caption{ Trends of output $f$-ratio for four given EE as a function of fiber
\# and position in the SparsePak array for 13 of SparsePak's
fibers. All fibers are fed with an f/6.3 input beam. The fiber number,
from 1 to 82, corresponds to a fiber's position in the slit and fiber
foot. Fiber \#1 is on the inside edge of the foot, where the radius of
curvature is smallest.  Fibers within one fiber-diameter from the
array edge are marked as solid triangles; fibers within 2-4
fiber-diameters from the array edge are marked as 4-pointed stars; all
other (interior) fibers are marked as open circles. (No fibers are on
the edge since there is a one-fiber buffer.) Note the strong,
first-order trend in increased FRD (smaller $f$-ratio at fixed EE) with
decreasing fiber number (foot radius), and the second-order trend with
distance from the edge of the fiber array.  Solid lines are WLS
regressions to all data (Akritas \& Bershady, 1996). Lower and
upper dashed lines are regressions to the outer ($|$dx$|=1$) and inner
($|$dx$|\geq2$) fibers, respectively; their offset from the solid line
at fiber \#42 is labeled. The $f$-ratio value for the reference cable is
marked as a horizontal dashed line.  SparsePak fibers approach this
limit when they are least bent and more than 2 fibers from the edge of
the array.  Substantial gains could be had by proper termination of
the fiber array cable, most particularly at the spectrograph end,
i.e., by straightening the foot.}
\end{figure}

\clearpage


\begin{figure}
\figurenum{9}
\plotfiddle{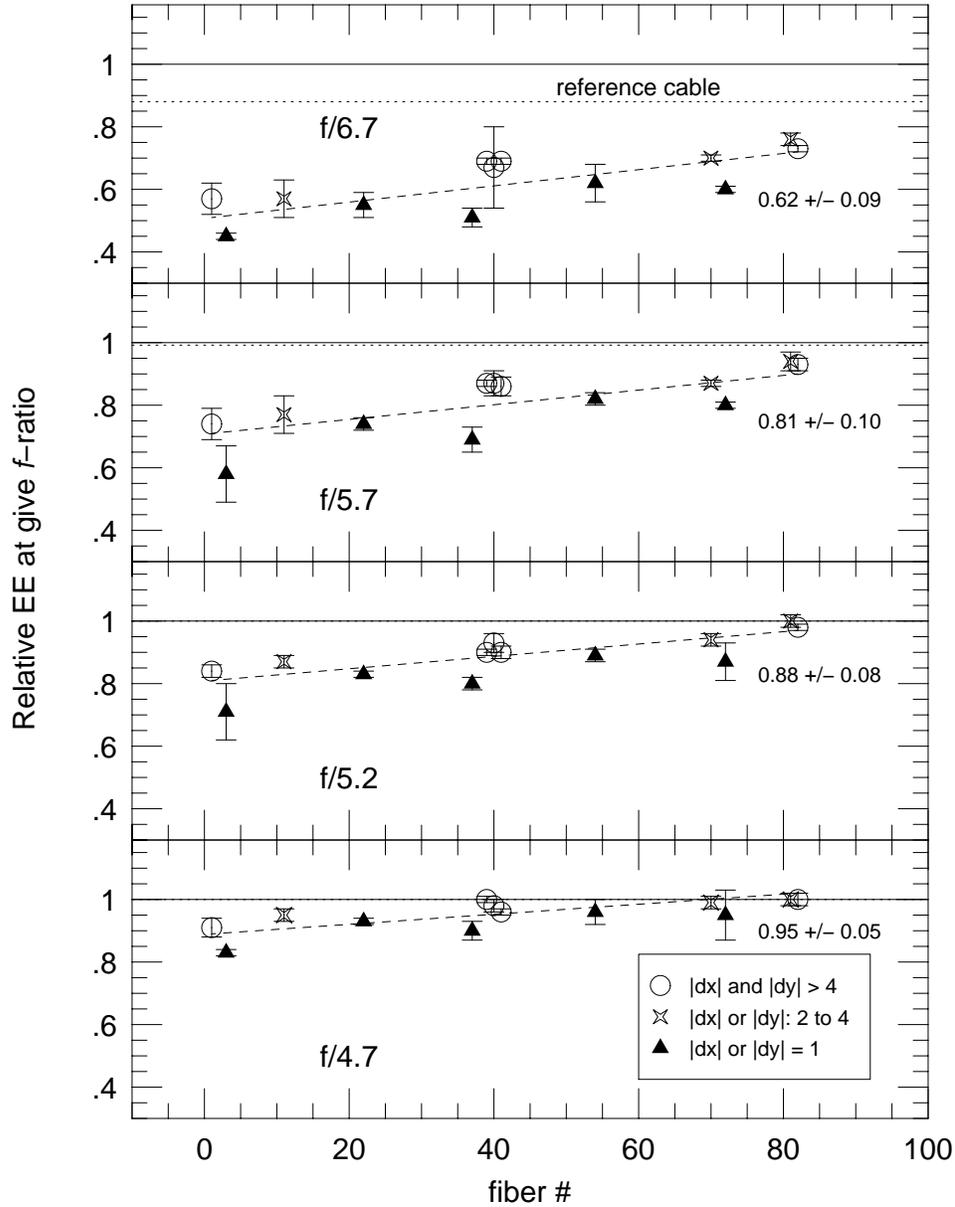}{6.0in}{0}{75}{75}{-240}{-50}
\caption{ Trends of relative EE vs fiber number for four output $f$-ratio:
6.7, 5.7, 5.2, and 4.7. Symbols are as in Figure 17 and key.  Dashed
lines are WLS regressions to all data (Akritas \& Bershady, 1996), and
are labeled for the mean fiber output EE for SparsePak. The dotted
line represents the output EE for the reference cable. The solid line
at unity is for reference.  The Bench Spectrograph optics is designed
for a 152mm collimated beam, which is achieved with the current
collimator in an f/6.7 beam. Hence, in the current configuration this
beam only contains between 45-75\% of the light (62\% on average).}
\end{figure}

\clearpage


\begin{figure}
\figurenum{10}
\plotfiddle{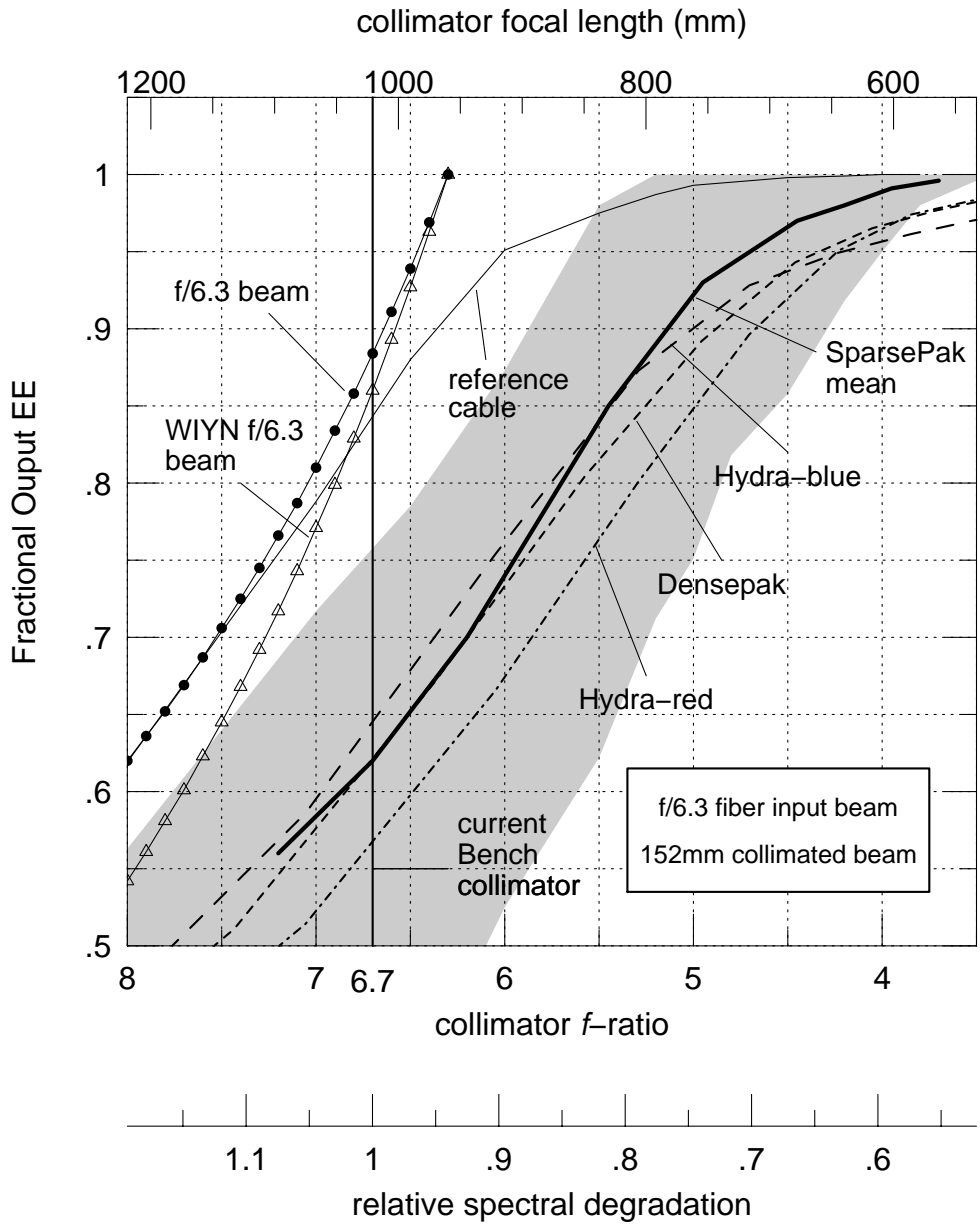}{6.0in}{0}{80}{80}{-240}{-50}
\caption{Fractional output encircled energy (EE) from fibers as a
function of WIYN Bench Spectrograph collimator focal ratio (bottom
axis) or collimator focal length (top axis). This assumes a 152mm
collimated beam diameter and that all fibers are fed with a f/6.3
beam. The mean SparsePak beam profile (thick, solid line) the range
for the 13 measured SparsePak fibers (grey shaded area), and reference
cable (thin solid curve) are based on laboratory measurements
(cf. Figure 15) using the f/6.3 input beam shown as a thin solid curve
with solid circles. Comparable curves, as measured {\it on the
telescope} for Densepak (300$\mu$m fibers), and the two Hydra cables
(with ``blue,'' 310$\mu$m and ``red,'' 200$\mu$m fibers), are shown
for comparison (private communication, P. Smith \& C. Conselice; see
text for further details). These measurements use the WIYN f/6.3 beam
(accounting for the central obstruction -- see text), shown as the
thin solid curve with open triangles. The very bottom scale (relative
spectral degradation) indicates how the spectral resolution of the
Bench would alter (worst case) due to changes in system
demagnification as a function of changes in the collimator focal
length at fixed camera focal length. The current Bench has a f/6.7
collimator for a 152mm collimated beam. This figure illustrates the
effects of FRD on light losses for the Bench Spectrograph, and how
optimization trades might be made between throughput and spectral
resolution for redesign of the Bench Spectrograph collimator.}
\end{figure}


\clearpage


\begin{figure}
\figurenum{11}
\plotone{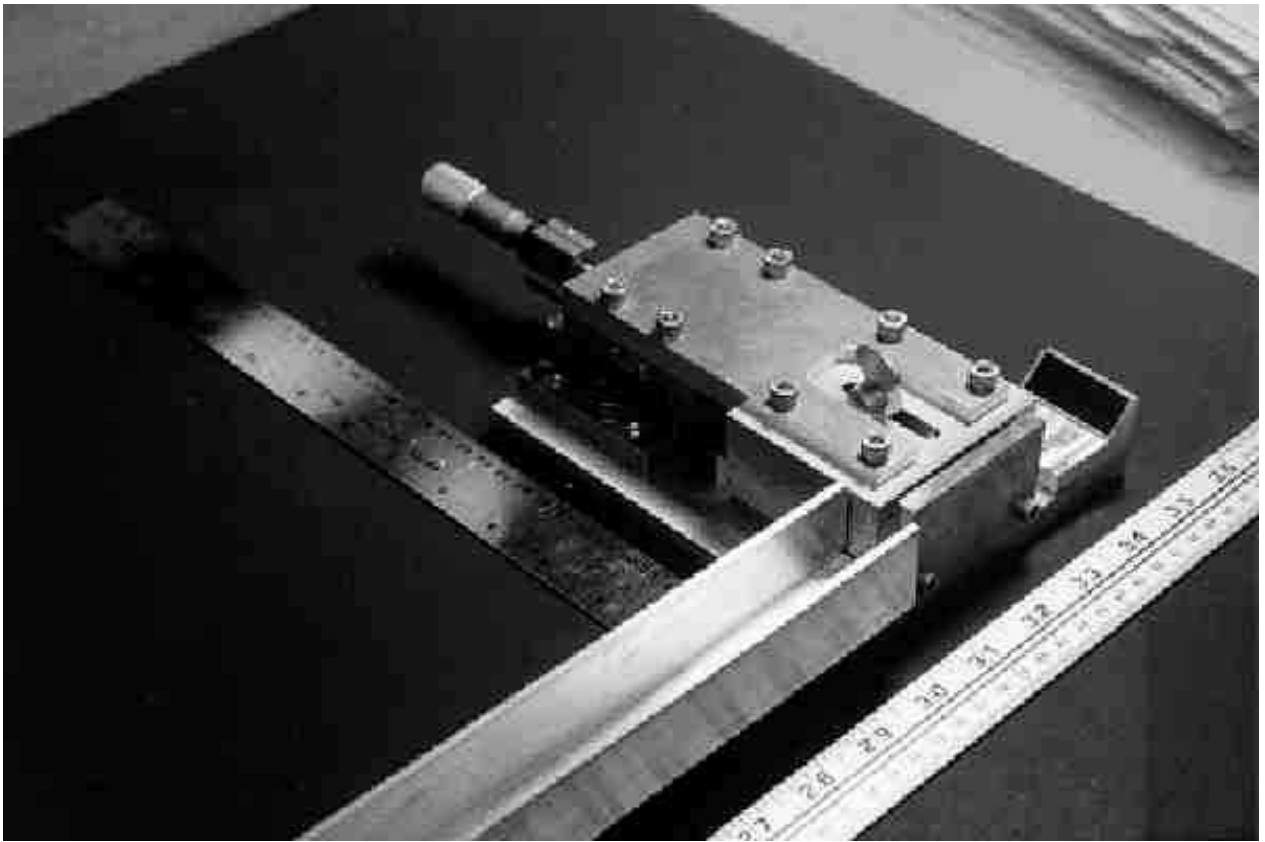}
\caption{ Pressing jig for gluing fiber arrays, based on a design by
S. Barden. The inverted, U-shaped pressing channel, lower right, has a
width controllable by a micrometer adjusted translation stage.  A
45$^\circ$ mirror (middle-right) is used for inspecting the fiber
array during construction and curing. The lower diagonal rail is for
holding the fiber cable, not used in practice. The entire jig is
designed to be taken apart and cleaned between presses.}
\end{figure}

\clearpage

%


\begin{figure}
\figurenum{12}
\plotfiddle{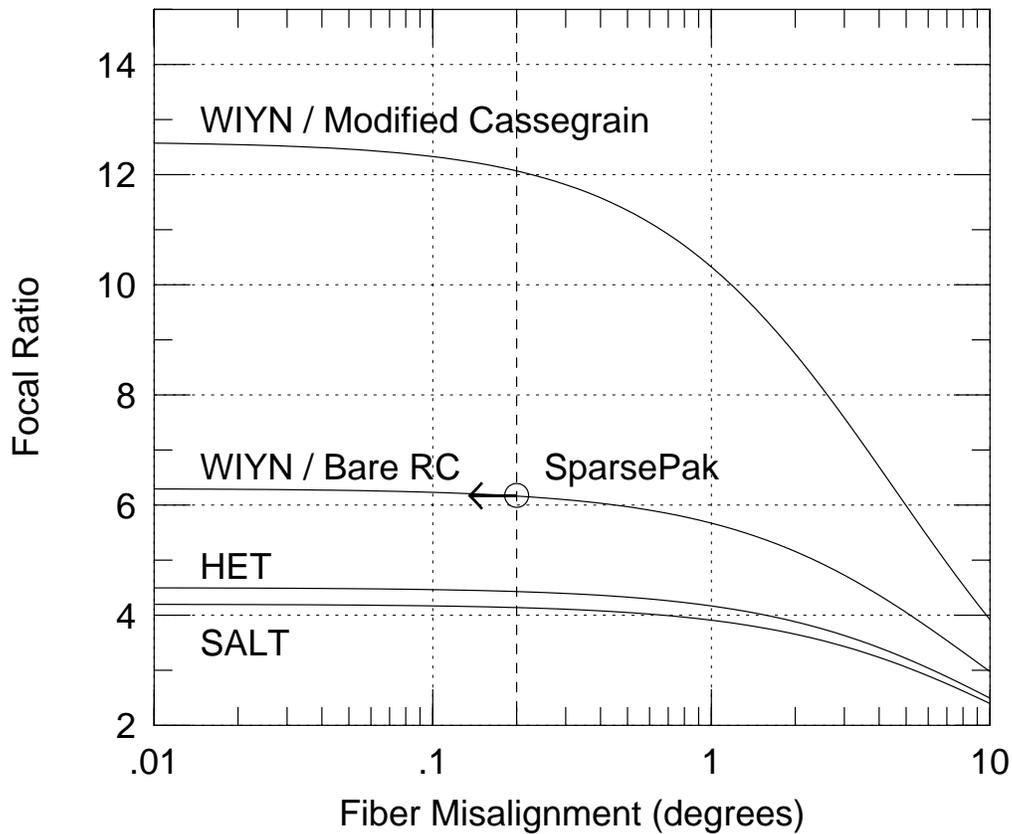}{5.0in}{0}{90}{90}{-280}{-125}
\caption{ Focal-ratio degradation (FRD) due to fiber mis-alignment with
the telecentric angle. The effective output $f$-ratio of the extreme
ray as a function of fiber misalignment from the optical axis (e.g.,
the telecentric angle) is shown for four input focal-ratios: (i) the
WIYN Modified Cassegrain port (CassIAS) at f/13.5; (ii) the WIYN
Nasmyth (bare RC) port (IAS) at f/6.3; (iii) the HET fiber instrument
feed (FIF) at f/4.5; and the SALT fiber instrument feed at f/4.2. The
effects are more pronounced for slower input beams, but in general
substantial degradation occurs for misalignments above
0.1-0.3$^\circ$. The estimated alignment error budget for SparsePak is
indicated by the vertical dashed line. In general, the effective
output $f$-ratio, as defined, e.g., by some encircled energy, will
also decrease with fiber misalignment, but at a slower rate depending
on the detailed shape of the input beam (see, for example, Wynne
1993).}
\end{figure}

\clearpage



\clearpage


\begin{figure}
\figurenum{13}
\epsscale{}
\plotone{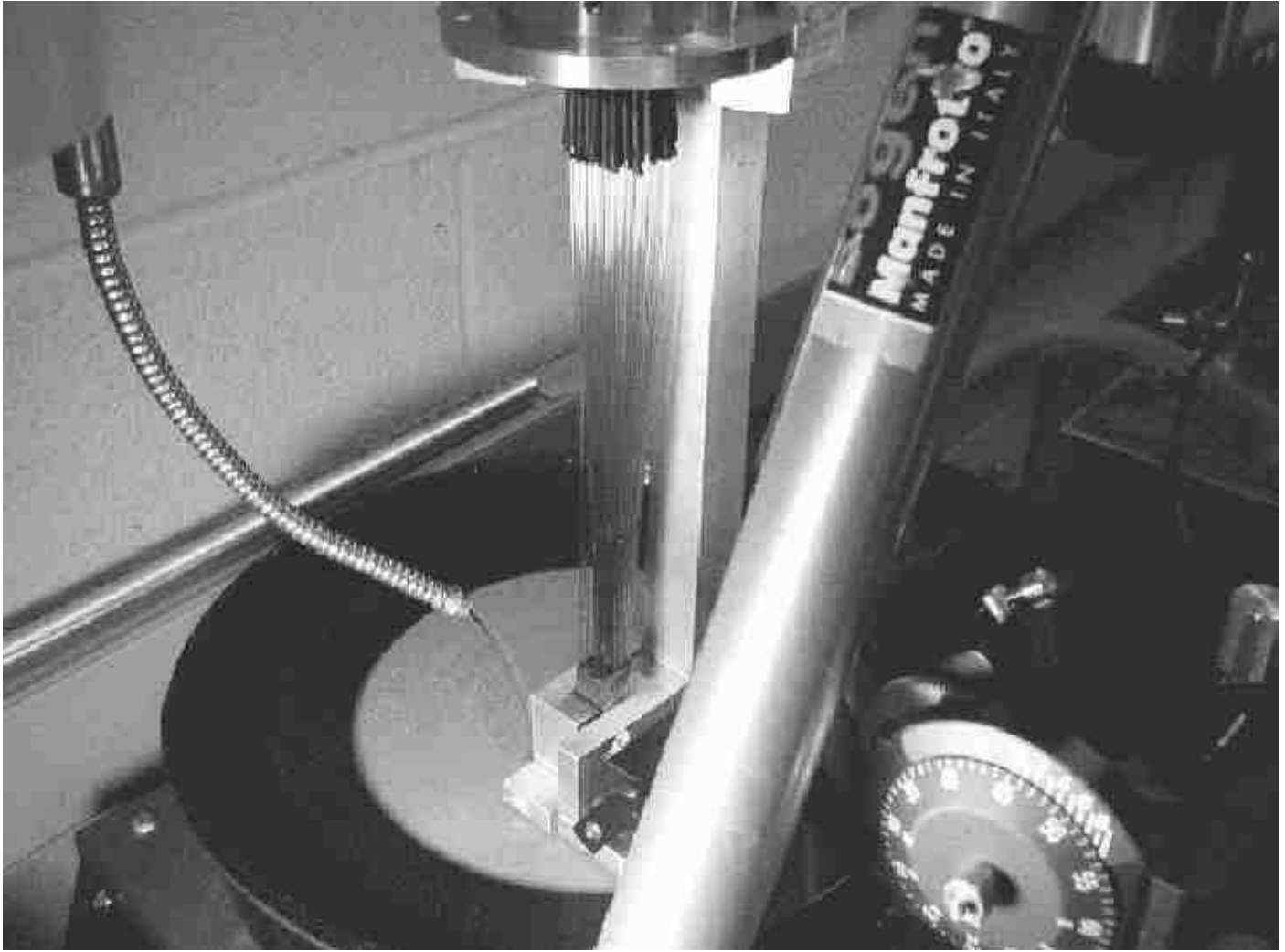}
\caption{ Close-up view of the SparsePak head polishing in progress on
the Ultrapol lap polisher. Note the vertical brace behind the bare
fibers which extends from the double L-bracket head attachment up to
the cable termination. One the L-brackets has padded walls which
serves to press the array against the other L-bracket to form a
precise, perpendicular alignment of the fiber array with the polishing
surface. The other L-bracket has U-shaped slot (not seen) for
attachment to the Ultrapol lapping polisher. The brace serves to
prevent compression of the fibers during polishing. Lubricant is
flowing through the flexible conduit onto the rotating platon
containing a polishing disk. Also visible to the lower right is the
Ultrapol polishing arm with precision angle (roll and pitch) and
height control. Given the load of the SparsePak cable, it was not
feasible to use this feature during this stage of the
polishing. Alignment of the fiber face perpendicular to the platen was
made relative to the L-bracket face. Note the leg of the camera
tripod, bearing the cable load. The cable is also supported by a
mechanical horse immediately to the left of the image.}
\end{figure}

\clearpage



\clearpage



\clearpage


\begin{figure}
\figurenum{14}
\plotone{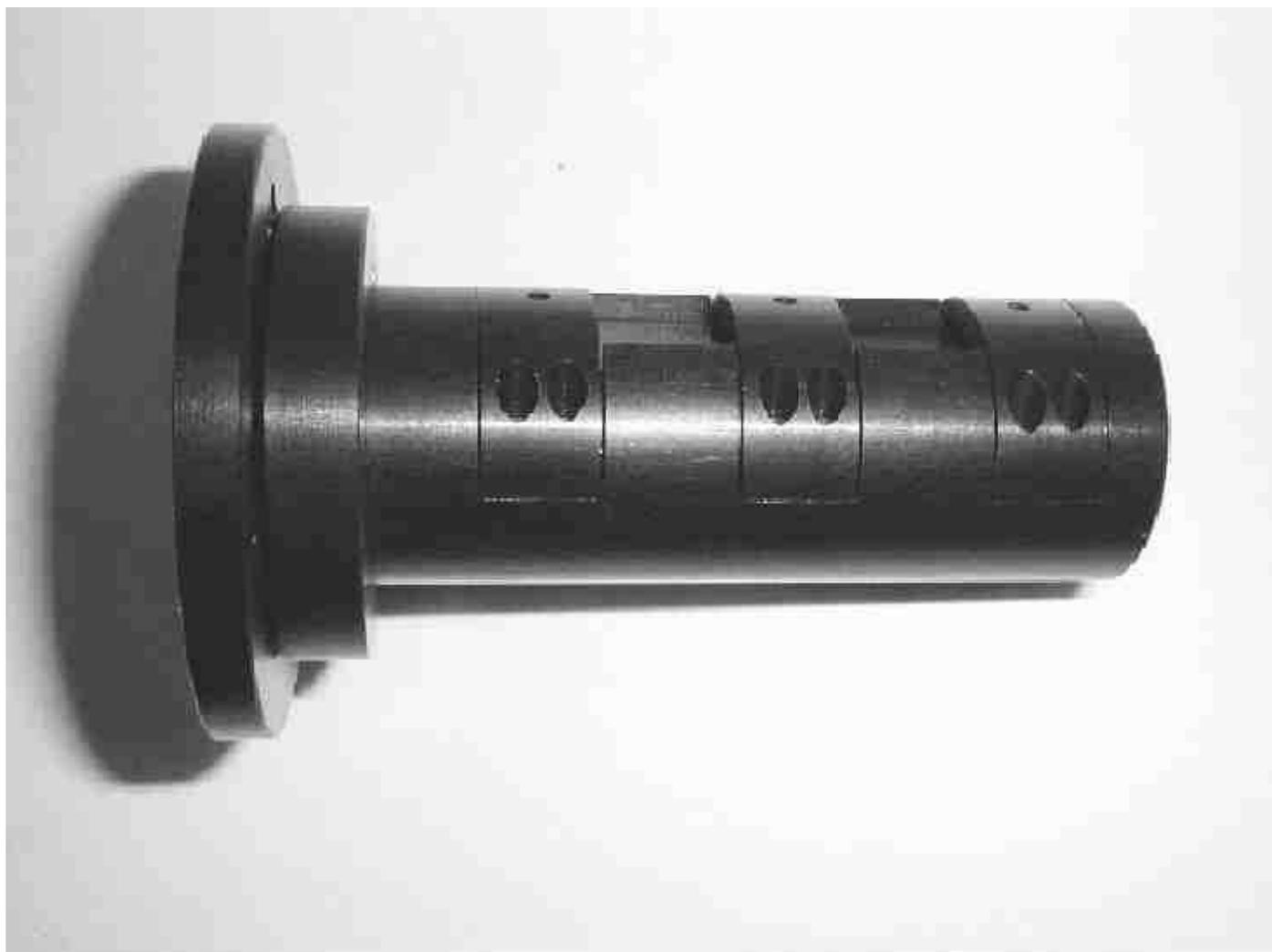}
\caption{ SparsePak fiber head-mount. The fiber bundle inserts from the
left, and terminates 7.6mm from the right end of the mount. Visible
are the three clam-shells, each with two pairs of mounting screws
(near pair visible), and a set-screw (top) which serves to press the
clamping-bar onto the array. The channel holding the array and bar is
visible at the top.}
\end{figure}

\clearpage


\begin{figure}
\figurenum{15}
\plotone{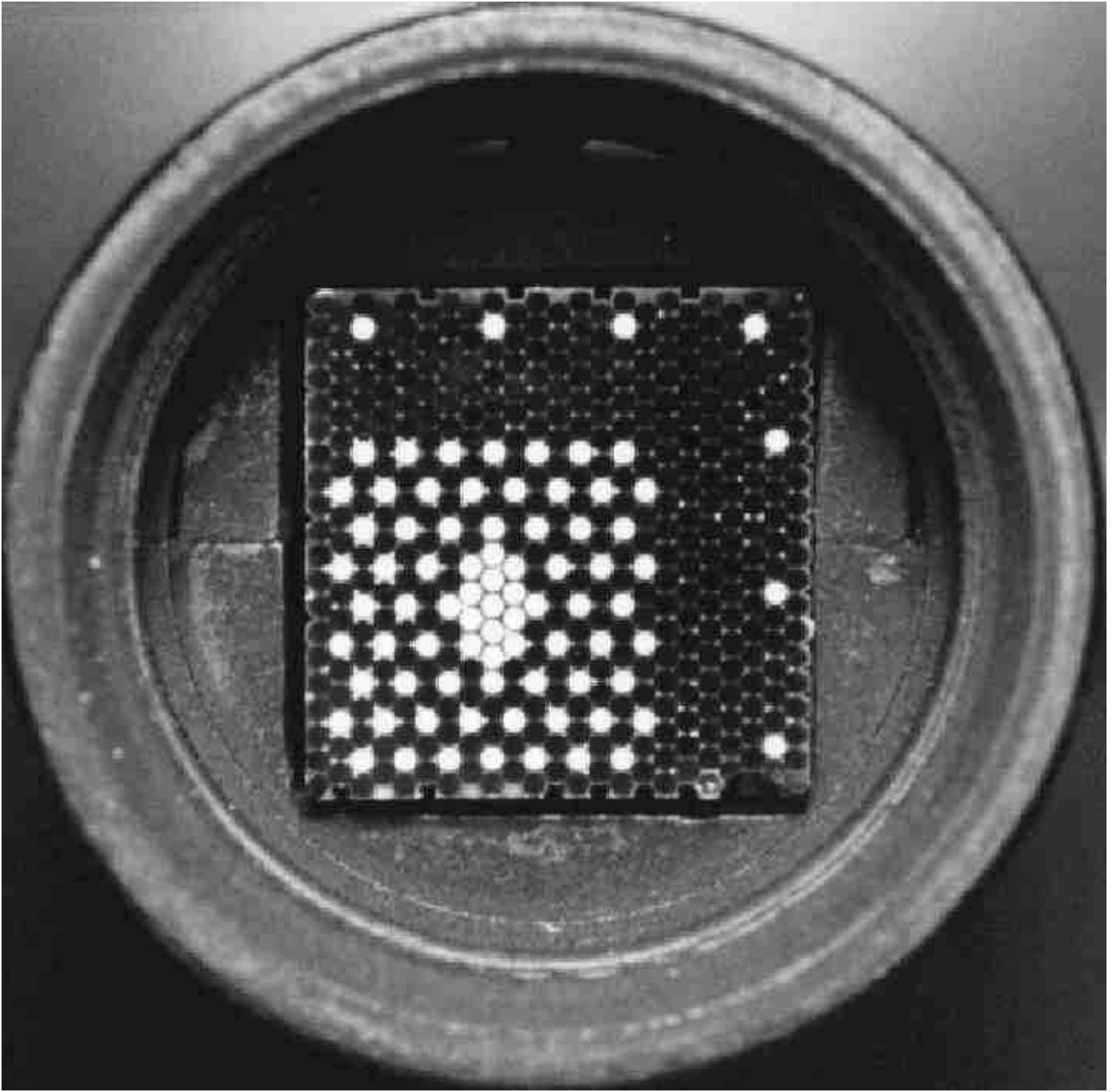}
\caption{ Final, polished SparsePak head in its assembled
head-mount. The end of the head-mount is beveled to avoid vignetting
the outermost fibers for input $f$-ratio slower than f/6; ridging and
flocking is to help eliminate scattered light. Back-illumination
indicates the active (light) and short, packing fibers (dark). On-sky
orientation at the WIYN IAS port with zero rotator offset places N
upward and E to the left (also as viewed in the WIFOE camera
slit-viewing camera).}
\end{figure}

\clearpage


\begin{figure}
\figurenum{16}
\plotone{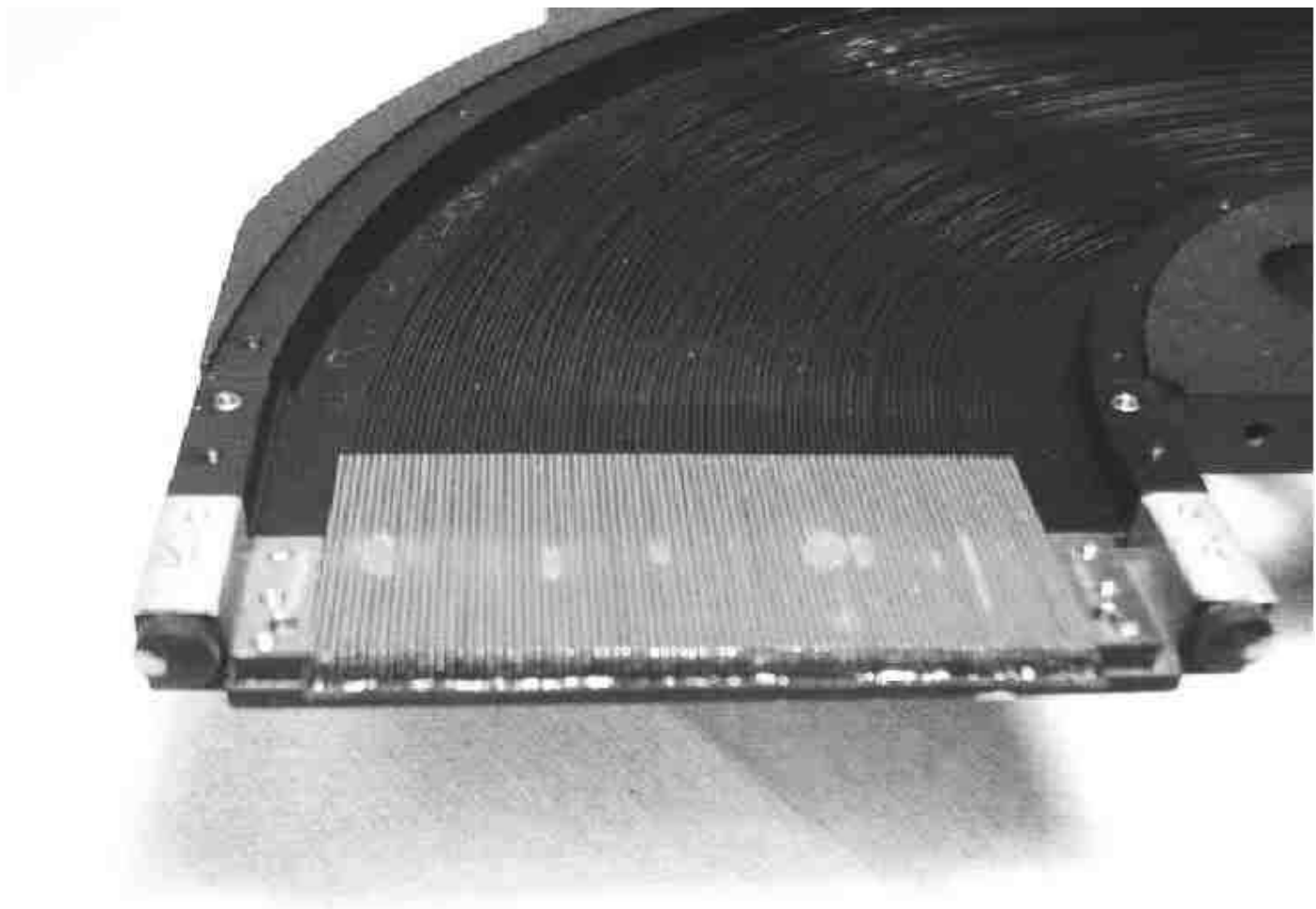}
\caption{ SparsePak fiber-foot and slit-block
mounted into its polishing attachment (non-anodized end). While only
the bottom portion of the fiber-foot is visible in this image, the
90$^\circ$ bend of the fibers is clearly visible. The 32mm-long
micro-tubes used for the slit block is at the bottom, with glued
fibers protruding several mm from the end. During polishing and final
assembly into the foot, a top clamp holds the slit-block in place. The
entire slit-block clamping assembly is pinned into the foot to ensure
precise location. The toes (not show here) extend from the end of the
slit block, toward the bottom and out of the page.}
\end{figure}

\vfil

\end{document}